\documentclass[twocolumn,aps,superscriptaddress,pra,longbibliography,floatfix]{revtex4}

\usepackage[multidot]{grffile}
\usepackage[normalem]{ulem}
\usepackage{amsmath,amssymb,graphicx}
\usepackage{mathtools, nccmath, textcomp}
\usepackage{subfigure}
\usepackage{algorithmic}
\usepackage{enumerate}
\usepackage{times}
\usepackage{color}
\usepackage{soul}
\definecolor{yblue}{rgb}{0.06, 0.3, 0.57}
\usepackage[pdftex]{hyperref}
\hypersetup{colorlinks=true,linkcolor=blue,citecolor=blue,urlcolor=blue}

\newcommand{\thintimes}{{\mkern-1mu\times\mkern-1mu}}

\begin{document}

\title{Existence, Stability and Dynamics of Monopole and
\\[0.25ex]
Alice Ring Solutions in Anti-Ferromagnetic Spinor Condensates}

\author{Thudiyangal Mithun}
\affiliation{Department of Mathematics and Statistics,
University of Massachusetts, Amherst MA 01003-4515, USA}

\author{R. Carretero-Gonz{\'a}lez}
\affiliation{Nonlinear Dynamical Systems
Group\footnote{\texttt{URL}: http://nlds.sdsu.edu},
Computational Sciences Research Center\footnote{\texttt{URL}: http://csrc.sdsu.edu/},
and Department of Mathematics and Statistics,
San Diego State University, San Diego, CA 92182-7720, USA}

\author{E.G. Charalampidis}
\affiliation{Mathematics Department,
California Polytechnic State University,
San Luis Obispo, CA 93407-0403, USA}

\author{D.S. Hall}
\affiliation{Department of Physics and Astronomy,
Amherst College,
Amherst, MA 01002-5000, USA}

\author{P.G. Kevrekidis}
\affiliation{Department of Mathematics and Statistics,
University of Massachusetts, Amherst MA 01003-4515, USA}

\begin{abstract}
In this work we study the existence, stability, and dynamics of
select topological point and line defects
in anti-ferromagnetic, polar phase, $F=1$ $^{23}$Na
spinor condensates. Specifically, we leverage fixed-point and
numerical continuation techniques in three spatial dimensions
to identify solution families of monopole and Alice rings as the
chemical potential (number of atoms) and
trapping strengths are varied within intervals of
realizable experimental parameters.
We are  able to follow the monopole from the linear limit of small atom number
all the way to the Thomas-Fermi regime of large atom number.
Additionally, and importantly, our studies reveal the existence of
{\it two} Alice ring solution branches,
corresponding to, relatively, smaller and larger ring radii, that
bifurcate from each other in a saddle-center bifurcation
as the chemical potential is varied.
We find that the monopole solution is always
dynamically unstable in the regimes considered.
In contrast, we find that the larger Alice ring is indeed stable close to
the bifurcation point until it destabilizes from an oscillatory
instability bubble for a larger value of the chemical potential.
We also report on the possibility of dramatically reducing,
yet not completely eliminating, the instability rates for the
smaller Alice ring by varying the trapping strengths.
The dynamical evolution of the different unstable waveforms is
also probed via direct numerical simulations.
\end{abstract}

\maketitle

%%%%%%%%%%%%%%%%%%%%%%%%%%%%%%%%%%%%%%%%%%%%%%%%
\section{Introduction}
%%%%%%%%%%%%%%%%%%%%%%%%%%%%%%%%%%%%%%%%%%%%%%%%

The study of Bose-Einstein condensates (BECs) is a topic that has now
enjoyed two and a half decades of substantial renewed interest given
the experimental (and theoretical) developments in the field that have
now been succinctly summarized in a number of
books~\cite{pethick,stringari}. While the simplest scenario of
such BECs has involved single-component/single-species atomic
gases, it was quickly realized that the internal degrees of freedom
of, e.g., different hyperfine states or the condensation of multiple
gases could offer a significant laboratory of
novel physical
phenomena including phase separation, spin dynamics, and domain
walls, among many others~\cite{stamper1998optical,stenger1998spin,stamper1999quantum,chang2005coherent,widera2006precision}. Such
multi-component systems are often studied in the realm
of so-called spinor condensates corresponding to hyperfine states
of $F=1$ and $F=2$ and, as such, have been the subject of numerous
dedicated
reviews~\cite{KAWAGUCHI2012253,Stamper_2013,kevrekidis2016solitons},
as well
as of books~\cite{pethick,stringari,frantzeskakis2015defocusing}.

A key feature of spinor condensates is the  presence of both
spin-independent and spin-dependent
interactions~\cite{KAWAGUCHI2012253,Stamper_2013}. Depending on the
sign of the latter, the gases can be anti-ferromagnetic,
as in the case of $^{23}$Na~\cite{stamper1998optical,stenger1998spin}
or ferromagnetic (weakly or strongly, respectively) for
$^{87}$Rb~\cite{chang2005coherent,widera2006precision} and
$^{7}$Li~\cite{huh2020observation,kim2021emission}.
The phase diagram of the potential ground states (depending on
the spin-dependent interactions and the role of external magnetic
fields through the so-called quadratic Zeeman effect) has been
extensively studied, e.g., in Ref.~\cite{KAWAGUCHI2012253}.

Beyond such ground states, part of the wealth of the phenomenology enabled by the
3-component $F=1$ and the 5-component $F=2$ BECs concerns
the possibility of solitonic (both topological and non-topological)
excitations in them~\cite{frantzeskakis2015defocusing}. Indeed,
at the one-dimensional (1D) level, both magnetic and non-magnetic structures
have been theoretically proposed and experimentally
observed~\cite{li2005exact,zhang2007solitons,nistazakis2008bright,szankowski2011surprising,romero2019controlled,chai2020magnetic,chai2021magnetic,bersano2018three,fujimoto2019flemish}
whence the phase diagram~\cite{katsimiga2021phase}, as well as
multi-soliton collisions~\cite{lannig2020collisions} of some of these
structures have been theoretically explored and experimentally quantified.
The presence of the spin degree of freedom has also enabled the observation
of spin domains~\cite{miesner1999observation,swislocki2012controlled}
and spin textures~\cite{ohmi1998bose,song2013ground}.
Although there are numerous topological and vortical structures
that have been uncovered in this
setting~\cite{al2001skyrmions,mizushima2002mermin,reijnders2004rotating,mizushima2002axisymmetric,leslie2009creation},
including even quantum knots~\cite{dsh2,dsh3}, our focus herein will
be on the existence, stability and dynamical properties
of elusive structures that have recently found a fertile ground
for their creation in this spinor setting, namely
monopoles~\cite{Stoof_2001,ray2015observation,dsh1}
and Alice rings~\cite{Ruostekoski_2003}.

To that end, we will first explore the monopole structure from a
dynamical systems point of view. Initially, we will explore its existence,
showcasing how it can be started in a three-dimensional (3D) parabolic confinement 
setting from the near-linear, low-atom-number limit, and continued towards
the large-atom-number, nonlinear (so-called Thomas-Fermi) regime.
We will subsequently develop stability diagnostics for this 3D
spatial structure as an equilibrium state and identify the modes that
render it unstable. The evolution of such an instability will offer
a hint towards the existence of a symmetry-broken state, in the form
of the so-called Alice ring (AR)~\cite{Ruostekoski_2003,PRA_DSH_MM}.
The latter state comes in two variants (with a smaller and a larger
ring, respectively) which terminate, via a saddle-center bifurcation,
at the same turning point, i.e., at a minimal value of the chemical
potential for which such solutions can exist.
The stability and dynamics of such ARs are examined as well.

Our presentation is structured as follows. In Sec.~\ref{sec:prelim}, we present
the general setup of the spinor model and discuss its main properties.
In Sec.~\ref{sec:monopole}, we present the features of the monopole solution,
while in Sec.~\ref{sec:ARs}, an analogous presentation is put forth for the AR.
Finally, in Sec.~\ref{sec:conclu}, we summarize our findings and present some
conclusions, as well as a number of directions for future  work.

%%%%%%%%%%%%%%%%%%%%%%%%%%%%%%%%%%%%%%%%%%%%%%%%
\section{Preliminaries}
\label{sec:prelim}
%%%%%%%%%%%%%%%%%%%%%%%%%%%%%%%%%%%%%%%%%%%%%%%%
\subsection{Model}
%%%%%%%%%%%%%%%%%%%%%%%%%%%%%%%%%%%%%%%%%%%%%%%%
The energy functional for the three spin-states wavefunction $\Psi=(\psi_{+1},\psi_0,\psi_{-1})$ of the
$F=1$ spinor condensate~\cite{KAWAGUCHI2012253,Robins_2001}
is given by
\begin{equation}
\label{eq:EnergyFunctional}
\mathcal{E}=\iiint_{-\infty}^{+\infty} d\mathbf{r}\Big\{\sum_{m=-1}^1 \psi_m^{\ast}\hat{h}_0\psi_m+\frac{g_0}{2}n^2+\frac{g_2}{2}|\mathbf{F}|^2\Big\},
\end{equation}
where the linear operator $\hat{h}_0=-\frac{\hbar^2}{2M}\nabla^2+V$ accounts for the kinetic energy
with $V=V(\textbf{r})$ corresponding to the external trapping potential for atoms of mass $M$.
The total density is defined as
\begin{align}
n=n_{-1}+n_0+n_{+1}=\sum_{m=-1}^1|\psi_m|^2,
\label{eq:totdens}
\end{align}
and the scattering coefficients are given by
$g_0={4\pi \hbar^2 a_s}/{m}$ and $g_2={4\pi \hbar^2 a_a}/{m}$
where $a_s$ and $a_a$ characterize, respectively, the spin-independent
and spin-dependent part of the interactions.
In particular, $a_s=(2 a_2 +a_0)/3$ and $a_a=(a_2-a_0)/3$~\cite{KAWAGUCHI2012253,Ohmi_1998_Bose,Ho_1998_spinor} with the
values of $a_0$ and $a_2$ given in Table 2 of Ref.~\cite{KAWAGUCHI2012253}.
The spin density vector $\mathbf{F}=(F_x,F_y,F_z)$ is given by
\begin{equation}
F_{a}=\sum_{m,n=-1}^1\psi_m^{\ast}(f_{a})_{mn}\psi_n,\quad a\in\{x,y,z\},
\label{spin_expec}
\end{equation}
where $f_{a}$ represent the Pauli's spin-1 matrices.
The components of the spin (magnetization) vector $\mathbf{F}$ are
\begin{eqnarray}
\notag
F_x&=&\frac{1}{\sqrt{2}}[\psi_{+1}^{\ast}\psi_0+\psi_0^{\ast}(\psi_{+1}+\psi_{-1})+\psi^{\ast}_{-1}\psi_0],\\
\notag
F_y&=&\frac{i}{\sqrt{2}}[-\psi_{+1}^{\ast}\psi_0+\psi_0^{\ast}(\psi_{+1}-\psi_{-1})+\psi^{\ast}_{-1}\psi_0],\\
\notag
F_z&=&|\psi_{+1}|^2-|\psi_{-1}|^2.
\end{eqnarray}
For the $F=1$ spinor condensate, the mean-field order parameter $\Psi$ can be expressed
as $\Psi(\mathbf{r})=\sqrt{n} e^{i\phi} \zeta(\mathbf{r})$, where $\phi(\mathbf{r})$
is the scalar global phase and $\zeta(\mathbf{r})$ represents the normalized spinor
which determines the average local spin $\langle \mathbf{F}\rangle = \mathbf{\zeta}^{\dagger} \mathbf{F}\mathbf{\zeta}$~\cite{Stoof_2001,Stamper_2013}.
The equations of motion for $\Psi$ ensuing from the energy
functional of Eq.~(\ref{eq:EnergyFunctional}) correspond to the following set of
spin-coupled Gross-Pitaevskii (GP) equations:
{\small
\begin{subequations}
\label{Eq:GPdim}
\begin{align}
 i\hbar \frac{\partial\psi_{+1}}{\partial t}&=\mathcal{H}\psi_{+1}+g_{2}(|\psi_0|^2+F_z)\psi_{+1}+g_2\psi_{-1}^{\ast}\psi_0^2,\\
 i\hbar \frac{\partial\psi_0}{\partial t}&=\mathcal{H}\psi_0+g_{2}(|\psi_{+1}|^2+|\psi_{-1}|^2)\psi_0+2g_2\psi_0^{\ast}\psi_{+1}\psi_{-1},\\
 i\hbar \frac{\partial\psi_{-1}}{\partial t}&=\mathcal{H}\psi_{-1}+g_{2}(|\psi_0|^2-F_z)\psi_{-1}+g_2\psi_{+1}^{\ast}\psi_0^2,
\end{align}
\end{subequations}
}
where
$$\mathcal{H}=\hat{h}_0+g_0\, n,$$
represents the spin-independent nonlinear operator, arising in the
famous
Manakov model~\cite{manakov1974}. The external potential in Eq.~\eqref{Eq:GPdim}
is assumed to be the same for all spin components, and is given by
$$V(\textbf{r})=\frac{1}{2}m\left( \Omega_x^2 x^2+ \Omega_y^2 y^2+\Omega_{z}^2 z^2\right),$$
where $\Omega_{a}$ is the trapping strength along the $a$th
direction. Note that we allow the parabolic trap to have independent strengths
along all spatial directions as we are also interested in studying the effects
on the existence and possible stabilization of the topological structures at hand,
namely monopoles and Alice rings (ARs), when introducing, relatively weak, in-plane $(x,y)$ anisotropies.

The spinor condensate admits multiple conserved quantities~\cite{KAWAGUCHI2012253}.
These are (i) the total atom number
\begin{equation}
\mathcal{N}=\mathcal{N}_{-1}+\mathcal{N}_0+\mathcal{N}_{+1},
\label{eq:mass}
\end{equation}
corresponding to the sum of all atom numbers for each component
(which are not individually conserved):
\begin{equation}
\label{eq:masseach}
\mathcal{N}_j=\iiint_{-\infty}^{+\infty} |\psi_j|^2\, dx\, dy\, dz, \quad j\in\{-1,0,+1\},
\end{equation}
(ii) the $z$-component of the magnetization
$$\mathcal{M}_z=\iiint_{-\infty}^{+\infty} F_z\, dx\, dy\, dz,$$
and (iii) the integral of the modulus squared of the magnetization
(spin) vector,
in addition to (iv) the energy of Eq.~(\ref{eq:EnergyFunctional})
which represents the Hamiltonian of the model.

For the simulations, we will resort to the dimensionless version of the GP equations
[cf. Eqs.~\eqref{Eq:GPdim}] by initially assuming a spherical trap with
$\Omega \equiv \Omega_x=\Omega_y=\Omega_{z}$ and  using the rescalings
$x \rightarrow a_0 x$, $t \rightarrow \Omega^{-1} t$, and
$\psi_j \rightarrow \sqrt{N} a_0^{-3/2} \psi_j$ where
$a_h=\sqrt{\frac{\hbar}{M \Omega}}$ is the harmonic oscillator length.
Note that, for ease of exposition, we keep the same notation as the
dimensional variables. However, from this point onwards
it is important to remember that all variables are dimensionless.
Under this adimensionalization, the GP equations become
\begin{equation}
\begin{split}
   \small i\frac{\partial\psi_{+1}}{\partial t}&=\mathcal{H}\psi_{+1}+c_{2}(|\psi_0|^2+F_z)\psi_{+1}+c_2\psi_{-1}^{\ast}\psi_0^2,\\
     \small i \frac{\partial\psi_0}{\partial t}&=\mathcal{H}\psi_0+c_{2}(|\psi_{+1}|^2+|\psi_{-1}|^2)\psi_0+2c_2\psi_0^{\ast}\psi_{+1}\psi_{-1},\\
       \small i\frac{\partial\psi_{-1}}{\partial t}&=\mathcal{H}\psi_{-1}+c_{2}(|\psi_0|^2-F_z)\psi_{-1}+c_2\psi_{+1}^{\ast}\psi_0^2,
     \end{split}
     \label{eq:GPd}
\end{equation}
where
\begin{eqnarray}
\mathcal{H}&=&-\frac{1}{2}\nabla^2+V(\textbf{r})+c_0\,n,\\[1.0ex]
V(\textbf{r})&=&\frac{1}{2}(x^2+y^2+z^2),
\end{eqnarray}
and $c_{0}=4\pi N a_{s}/ a_{h}$, while $c_{2}=4\pi N a_{a}/ a_{h}$.
Typically, $a_h/a_s$ varies between 500 and 4000~\cite{Leanhardt_2003_Lean,Ruostekoski_2003}.
Following Refs.~\cite{stenger1998spin,Black2007Spinor,Samuelis2000Cold},
we consider, in particular, a $^{23}$Na condensate with $a_a=(a_2-a_0)/3\approx 2 a_B$
and $a_s=(2 a_2 +a_0)/3\approx 50 a_B$ in our simulations where $a_B= 0.0529\,$nm is the
Bohr radius, thus giving $c_2/c_0\approx 0.04$. For instance, for trapping frequencies
$\Omega=2\pi\times10\,$Hz and and $N\approx 4\times 10^6$, we get $c_0=2\times 10^4$.

%%%%%%%%%%%%%%%%%%%%%%%%%%%%%%%%%%%%%%%%%%%%%%%%%%%%%%%%%
\subsection{Ground State Phases}
%%%%%%%%%%%%%%%%%%%%%%%%%%%%%%%%%%%%%%%%%%%%%%%%%%%%%%%%%

In the absence of magnetic field (Zeeman terms), the ground state can
be either (i) ferromagnetic (${\cal M}_z \neq 0$ and $c_2<0$) or
(ii) anti-ferromagnetic or polar (${\cal M}_z=0$ and $c_2>0$)~\cite{KAWAGUCHI2012253}.
In the present work, we focus on the polar case where the order
parameter is $(0,1,0)^T$, i.e., the hyperfine state corresponding
to $m_f=0$ is the only one populated at the ground state.
For a $F=1$ spinor condensate, the mass velocity acquires the form
\begin{equation}\label{massv}
 \mathbf{v}^{\rm mass} = \frac{\hbar}{M}(\nabla \phi-i\mathbf{\zeta}^{\dagger} \nabla \mathbf{\zeta}).
\end{equation}
Nevertheless, the absence of magnetization in the polar phase simplifies
the mass velocity to $\mathbf{v}^{\rm mass} = \left(\hbar/M\right)\nabla \phi$.
As a result, mass circulation along a loop is quantized; yet, the
single-valuedness of the order parameter, $\Psi$, demands the
quantization of
circulation
in units of $h/(2M)$ (half-quantum vortex)~\cite{KAWAGUCHI2012253}. This
stems from the fact that for ${\cal M}_z=0$, the order parameter $\Psi$ can be
represented as
\begin{equation}
 \Psi=\sqrt{n}e^{i\phi}\hat{\mathbf{d}},
 \label{eq:opm}
\end{equation}
where $\hat{\mathbf{d}}$ is a unit
vector that defines the quantization axis, and
that would yield ($\hat{\mathbf{d}}, \phi) \rightarrow (-\hat{\mathbf{d}}, \phi+\pi)$.
This invariance  reflects the $\mathbb{Z}_2$ symmetry of $\Psi$.

%%%%%%%%%%%%%%%%%%%%%%%%%%%%%%%%%%%%%%%%%%%%%%%%%%%%%%%%%%%%%%%
\section{Monopole}
\label{sec:monopole}
%%%%%%%%%%%%%%%%%%%%%%%%%%%%%%%%%%%%%%%%%%%%%%%%%%%%%%%%%%%%%%%

A monopole is a topologically excited point defect and is
characterized by the second homotopy group
($\pi_2$)~\cite{KAWAGUCHI2012253}.
It can be realized in a BEC either
by breaking the global symmetry of the condensate or by a
gauge-potential~\cite{Pietila_2009,KAWAGUCHI2012253}. These structures
can be naturally realized in a spinor condensate.
Indeed, since in the polar phase of the $F=1$ spinor condensate we are free to
change the (global) phase and spin, a monopole can be created by choosing the
(radial) hedgehog field
$\hat{\mathbf{d}}={\mathbf{r}}/{r}$ (where $r=|\mathbf{r}|$)
in Eq.~(\ref{eq:opm}) with
$\phi=0$ by minimizing the gradient energy~\cite{Ruostekoski_2003,Stoof_2001}.
Such a monopole is called
't-Hooft-Polyakov monopole~\cite{Ruostekoski_2003}.
The $\psi_{\pm 1}$ components
in the case of such a monopole feature
overlapped vortex lines
each carrying a single quantum of angular
momentum (but opposite circulation between the two).
On the other hand, $\psi_0$ resembles a planar dark
soliton (i.e., featuring a $\pi$-phase jump across the soliton plane)
as indicated in Ref.~\cite{Ruostekoski_2003}.
Moreover, for a monopole,
Eq.~(\ref{eq:opm}) vanishes at the origin, i.e., the density $n(\mathbf{0})=0$.

%%%%%%%%%%%%%%%%%%%%%%%%%%%%%%%%%%%%%%%%%%%%%%%%%%%%%%%%%%%%%%%
\begin{figure}[!pt]
 \includegraphics[width=0.99\columnwidth]{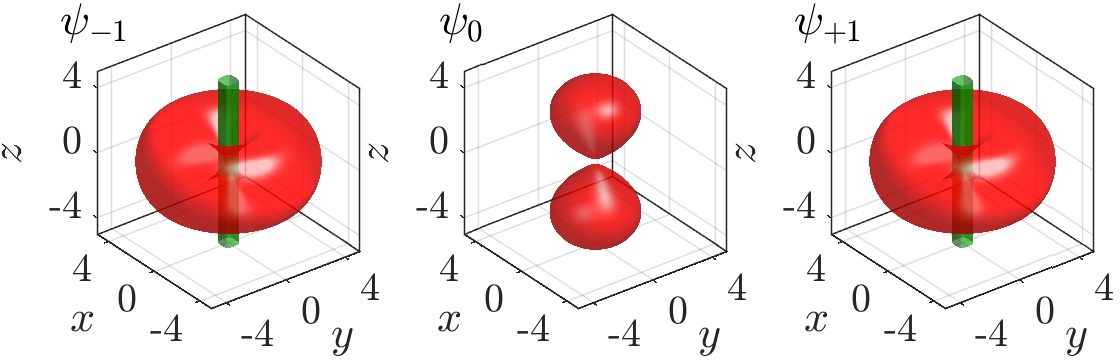}
\\[3.0ex]
 \includegraphics[width=0.80\columnwidth]{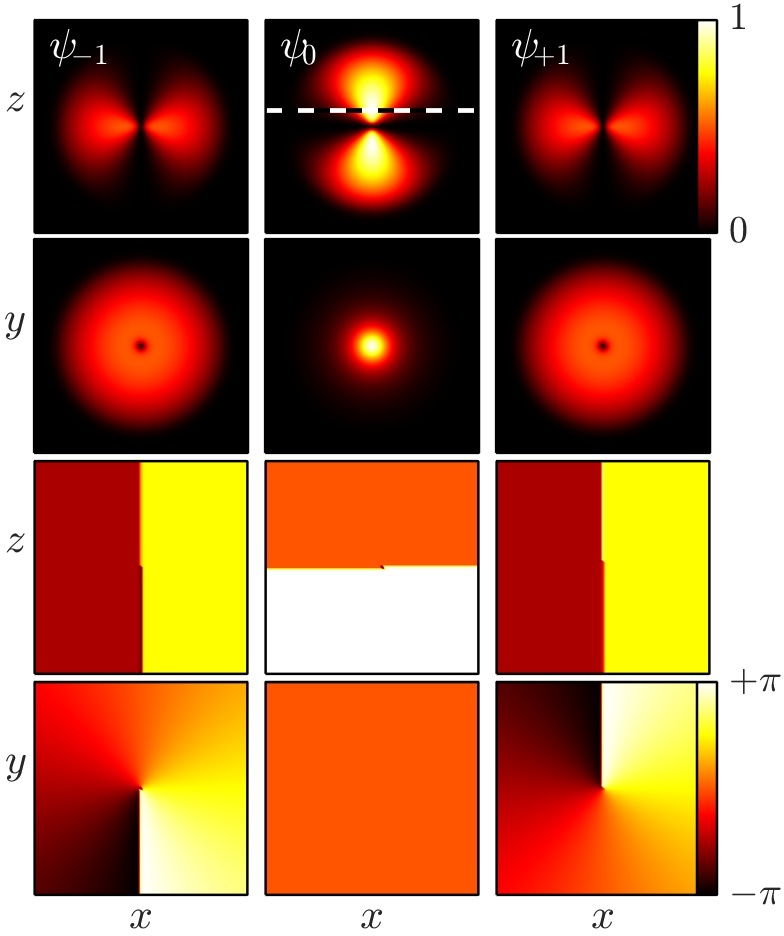}
 \caption{
\label{fig:monopole_mu20}
(Color online)
Monopole solution for $\mu=20$ in an isotropic trap with $\Omega=1$.
The top row of panels depicts isolevel cuts of constant density (red) and
vorticity (green) for the components $\psi_{+1}$ (left), $\psi_{0}$ (middle)
and $\psi_{-1}$ (right).
Second and third (fourth and fifth) rows depict cuts of the respective
densities (phases). All cuts pass through the origin except for the $(x,y)$ cuts
for $\psi_0$ which are done through the plane of maximum density shown
by the horizontal dashed line in the middle panel.
For ease of presentation, the depicted densities have been normalized so
that max($|\psi_{+1}|^2,|\psi_{0}|^2,|\psi_{-1}|^2)=1$.
The field of view for the cuts corresponds to
$-7.5<x,y,z<7.5$ while the computational domain is
$(x,y,z)\in[-L_x,L_x]\times[-L_y,L_y]\times[-L_z,L_z]$
where $L_x=L_y=L_z=12$.
}
\end{figure}
%%%%%%%%%%%%%%%%%%%%%%%%%%%%%%%%%%%%%%%%%%%%%%%%%%%%%%%%%%%%%%%

%%%%%%%%%%%%%%%%%%%%%%%%%%%%%%%%%%%%%%%%%%%%%%%%%%%%%%%%%%%%%%%
\begin{figure}[!htbp]
 \includegraphics[width=0.99\columnwidth]{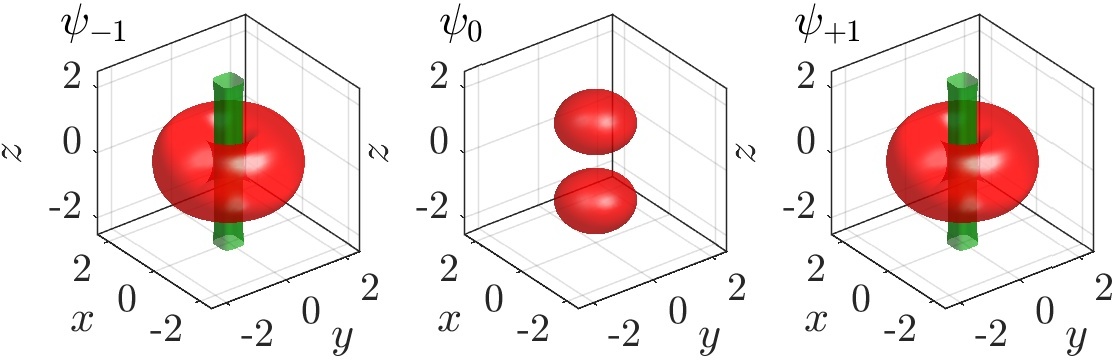}
\\[3.0ex]
 \includegraphics[width=0.80\columnwidth]{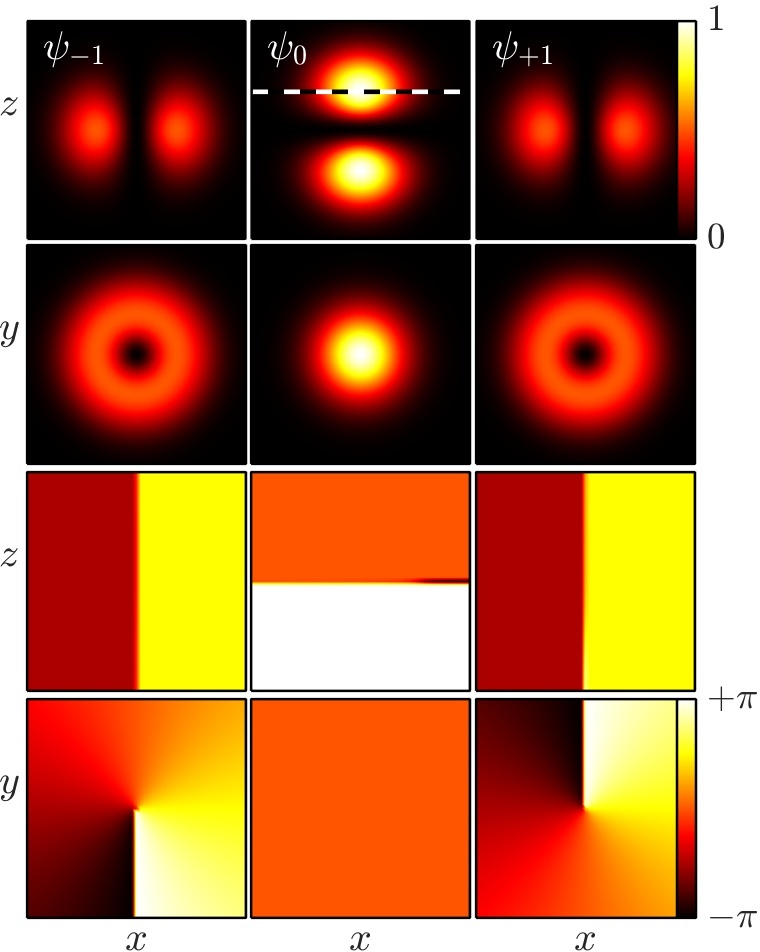}
 \caption{
\label{fig:monopole_mu3p5}
(Color online)
Same as in Fig.~\ref{fig:monopole_mu20} but for a monopole solution for $\mu=3.5$, i.e., 
close to the linear limit ($\mu=5/2$) which corresponds to a much lower atom number.
The field of view for the cuts corresponds to $-3<x,y,z<3$.
}
\end{figure}
%%%%%%%%%%%%%%%%%%%%%%%%%%%%%%%%%%%%%%%%%%%%%%%%%%%%%%%%%%%%%%%

%%%%%%%%%%%%%%%%%%%%%%%%%%%%%%%%%%%%%%%%%%%%%%%%%%%%%%%%%%%%%%%
\subsection{Steady States}
%%%%%%%%%%%%%%%%%%%%%%%%%%%%%%%%%%%%%%%%%%%%%%%%%%%%%%%%%%%%%%%

\begin{figure}[!htbp]
 \includegraphics[width=0.99\columnwidth]{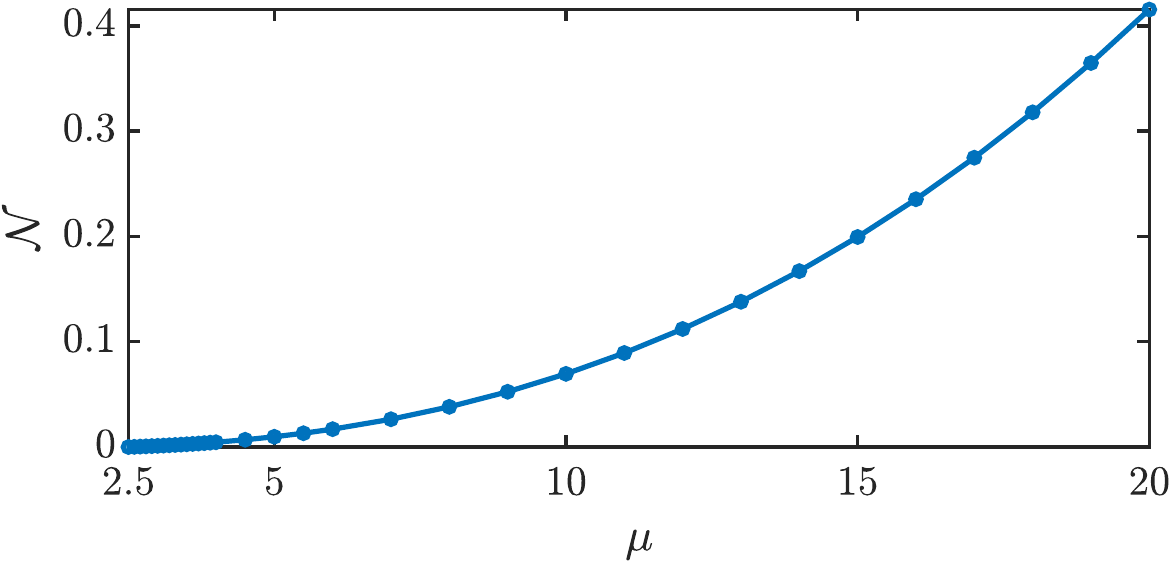}
\caption{
\label{fig:monopole_mass}
(Color online)
Total (adimensionalized) atom number [cf. Eq.~\eqref{eq:mass}] for the monopole
solution as the (adimensionalized) chemical potential $\mu$ is varied.
The monopole solution starts, in the linear limit, at $\mu= 5/2$.
}
\end{figure}

Steady states for the spinor BEC are obtained by considering stationary
solutions of the form $\psi_j(\textbf{r},t)\rightarrow\psi_j(\textbf{r})e^{-i\mu_j t}$
in Eq.~(\ref{eq:GPd}) for fixed (non-dimensional) chemical potentials $\mu_j$.
From these equations and direct substitution, it can be immediately
inferred that steady states are only possible if $\mu_{-1} + \mu_{+1}=2 \mu_0$. Considering a
symmetric split between the $\psi_{+1}$ and $\psi_{-1}$ components
(i.e., $\mu_{+1}=\mu_{-1}$), implies that
non-trivial configurations satisfy
\begin{equation}
\begin{split}
\label{eq:GPdSS}
(\mathcal{H}-\mu)\psi_{+1}+c_{2}(|\psi_0|^2+F_z)\psi_{+1}+c_2\psi_{-1}^{\ast}\psi_0^2&=0,\\
(\mathcal{H}-\mu)\psi_0+c_{2}(|\psi_{+1}|^2+|\psi_{-1}|^2)\psi_0+2c_2\psi_0^{\ast}\psi_{+1}\psi_{-1}&=0,\\
(\mathcal{H}-\mu)\psi_{-1}+c_{2}(|\psi_0|^2-F_z)\psi_{-1}+c_2\psi_{+1}^{\ast}\psi_0^2&=0,
\end{split}
\end{equation}
where $\mu\equiv\mu_{-1}=\mu_{+1}=\mu_0$. Note that different values of $\mu$ will
correspond to different total masses in the original system.
Steady states are then computed using
discretization methods and iterative solvers of the resulting coupled, nonlinear
algebraic equations
(see Appendix \ref{app:numA} for more details).
The initial seed 
is chosen to be a combination
of a dark soliton
(i.e., a quasi 2D planar structure) at $z=0$ in $\psi_0$ and vortex lines
of charge $\pm1$ in components $\psi_{\pm1}$, and given by
\begin{equation}
  \Psi=\begin{bmatrix}
    \psi_{+1}   \\
    \psi_0 \\
    \psi_{-1}
  \end{bmatrix}
 = \frac{\sqrt{n(\mathbf{r})} e^{i\phi} }{\sqrt{2}}\begin{bmatrix}
    -r_x+ir_y   \\
    \sqrt{2}r_z \\
    r_x+ir_y
  \end{bmatrix},
\end{equation}
with
$\hat{\mathbf{r}}=(r_x,r_y,r_z)=(x,y,z)/r$.
Without loss of generality, we fix $\phi=0$, and we note that
all components are
modulated by their respective Thomas-Fermi (TF) density
approximations, which,
assuming large density (and chemical potential), uses
\begin{equation}
 n(\mathbf{r})={\frac{\mu -V(\mathbf{r})}{c_0}},
 \label{eq:TF}
\end{equation}
when $\mu \geq V(\mathbf{r})$ and $0$ otherwise.
In Figs.~\ref{fig:monopole_mu20} and Fig.~\ref{fig:monopole_mu3p5} we depict the 
converged steady-state solutions for $\mu=20$ and $\mu=3.5$, respectively.
As the phase cuts and isolevel vorticity cuts depict, the $\psi_{+1}$ and $\psi_{-1}$
components contain a straight, vertical vortex line about the $z$-axis.
Also, the mutual repulsion between the $\psi_{\pm 1}$ and $\psi_0$
components,
along with the presence of the vortex line along the $z$-axis result
in the $\psi_{\pm 1}$ densities being pushed in the plane $(x,y)$ radially out while the
$\psi_0$ component splits into two domains along the $z$-axis separating regions of
opposite phase (see respective phase cuts).
The $\mu=3.5$ case corresponds to a configuration with small total
mass (i.e., atom number)  close to the linear limit ($\mu=5/2$; see below) of
the governing equations, whereby ${\cal N} \rightarrow 0$.
Comparing Figs.~\ref{fig:monopole_mu20} and \ref{fig:monopole_mu3p5}, it is evident
that the width of the vortex line for $\mu=3.5$ is significantly larger than the
corresponding one for $\mu=20$
as the corresponding healing length is inversely proportional to the
square root of the chemical potential.
Indeed, at the linear limit the vortex lines along the $z$-axis result from suitable
linear combinations of the harmonic oscillator modes $|p,l,k \rangle$,
corresponding to quantum numbers $p$, $l$, and $k$ along the
three Cartesian directions. These vortex lines with topological
charge $S= \pm 1$ arise from linear
combinations $|1,0,0 \rangle \pm i |0,1,0 \rangle$ (for the
components $\psi_{\pm 1}$) or similar, and the
component $\psi_0$ features a dipolar state $|0,0,1 \rangle$. All of
these linear states are energetically degenerate at the linear limit with
$p+l+k=1$ and hence correspond to energy
$E= \Omega (p+l+k + 3/2)=5/2$, as numerically
identified above. This, in turn, suggests that the monopole state is a
mode that directly emerges from the (first excited state within the) linear limit 
of the parabolically confined system, as showcased in Fig.~\ref{fig:monopole_mu3p5} 
and exists all the way to the highly nonlinear (TF)
limit thereof, as suggested in Fig.~\ref{fig:monopole_mu20}.
The dependence of the corresponding atom number on the chemical
potential in the path between these two limits can be visualized in
Fig.~\ref{fig:monopole_mass}.

%%%%%%%%%%%%%%%%%%%%%%%%%%%%%%%%%%%%%%%%%%%%%%%%%%%%%%%%%%%%%%%
%%%%%%%%%%%%%%%%%%%%%%%%%%%%%%%%%%%%%%%%%%%%%%%%%%%%%%%%%%%%%%%

%%%%%%%%%%%%%%%%%%%%%%%%%%%%%%%%%%%%%%%%%%%%%%%%%%%%%%%%%%%%%%%
\begin{figure}[!htbp]
 \includegraphics[width=0.99\columnwidth]{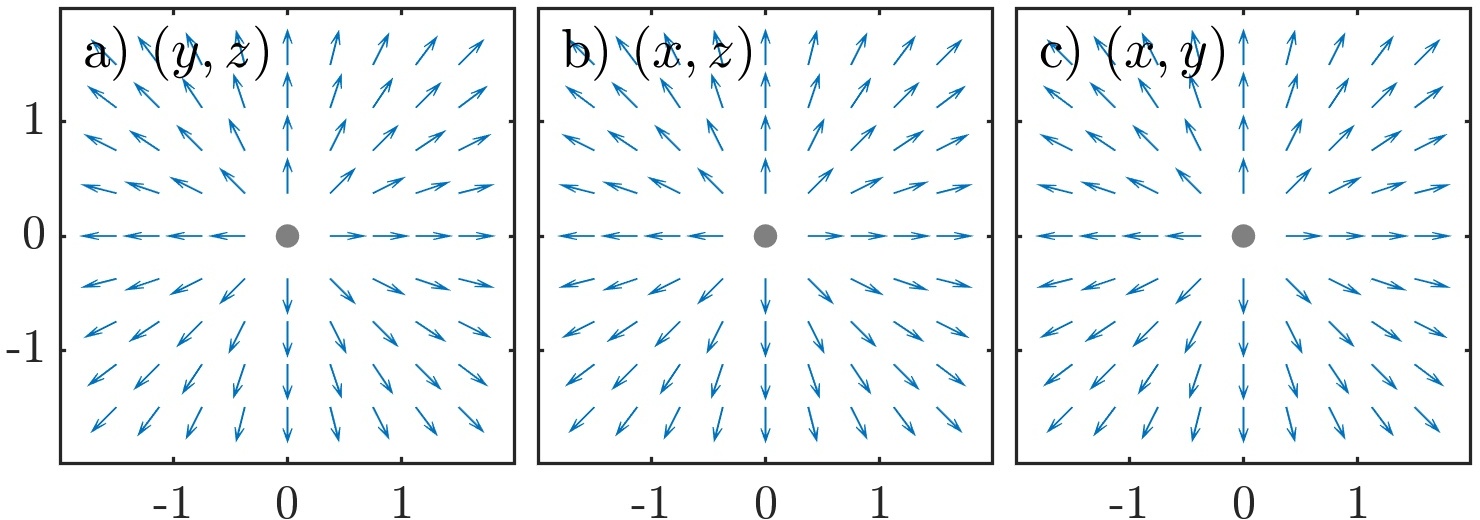}
\caption{
\label{fig:monopole_nematic}
Projections along the $(y,z)$ (left), $(x,z)$ (middle) and $(x,y)$
(right) plane of the nematic vector for the monopole solution for $\mu=20$.
The unequivocal hedgehog topology of the monopole (with the field
pointing radially outwards) can be observed.
}
\end{figure}
%%%%%%%%%%%%%%%%%%%%%%%%%%%%%%%%%%%%%%%%%%%%%%%%%%%%%%%%%%%%%%%

In order to showcase the monopole nature of the converged solutions,
we compute the nematic vector field (also referred to as the director field).
For a pure polar state (like the monopole
solution), the nematic vector field can be computed by decomposing the spinor
wavefunction through Eq.~\eqref{eq:opm}
where ${\mathbf{\hat{d}}}=(d_x,d_y,d_z)^T$ is in general known as the
nematic vector~\cite{PRA_DSH_MM}.  However,
for general solutions
with non-vanishing magnetization (as it will
be the case of the AR solutions presented in the next section), the above
description cannot be used.
Instead, we rely on the method of nematic vector field calculation through the
magnetic quadrupolar tensor~\cite{PRA_Muller,BlakieSymesPRA}
\begin{equation}
  \label{eq:Q}
{\mathcal Q}_{ab}=\frac{\zeta_a\zeta_b^*+\zeta_b\zeta_a^*}{2},
\end{equation}
where the $\zeta_j$ ($j\in\{x,y,z\}$) are the normalized spinor components expressed in
Cartesian coordinates~\cite{KAWAGUCHI2012253}, namely $\psi_j=\sqrt{n}e^{i\phi}\zeta_j$,
and
\begin{equation}
  \label{eq:SpinCart}
\left\{
\begin{array}{rcl}
\psi_{x} &=&\displaystyle \frac{\psi_{+1}-\psi_{-1} }{\sqrt{2}},
\\[1.5ex]
\psi_{y} &=&\displaystyle i\frac{\psi_{+1}+\psi_{-1} }{\sqrt{2}},
\\[1.5ex]
\psi_{z} &=&\displaystyle \psi_{0}.
\end{array}
\right.
\end{equation}
Then, following the prescription of, e.g., Ref.~\cite{PRA_DSH_MM},
one can extract the nematic vector by choosing, at each spatial location,
the eigenvector of ${\mathcal Q}$ corresponding to the largest eigenvalue
which corresponds to the local spin orientation of the order parameter.
According to this method to compute the nematic vector,
it should be noted that the factorization of the density and the global phase
$\phi$ are irrelevant [the density is a common scalar factor of ${\mathcal Q}$ 
and the phase $\phi$ gets eliminated through the combinations such as
$\psi_a\psi_b^*$ appearing in Eq.~(\ref{eq:Q})].
Figure~\ref{fig:monopole_nematic} depicts projections of the nematic vector
along different planes of the relevant 3D structure for a typical monopole.
As the figure shows, the nematic vector corresponds to a purely radial,
outward field evidencing the monopole texture of the solution.

%%%%%%%%%%%%%%%%%%%%%%%%%%%%%%%%%%%%%%%%%%%%%%%%%%%%%%%%%%%%%%%
\begin{figure}[!htbp]
\center{
\hspace{-0.5cm}
 \includegraphics[width=0.99\columnwidth]{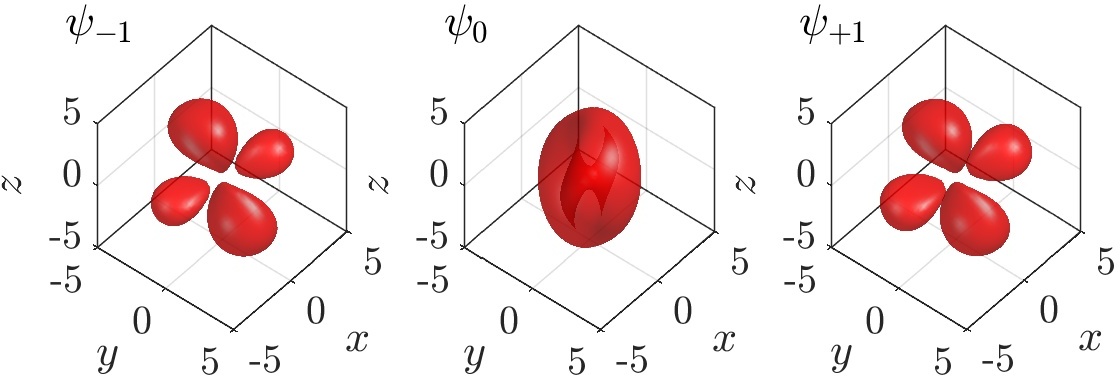}
\\[2.0ex]
\hspace{0.5cm}
 \includegraphics[width=0.80\columnwidth]{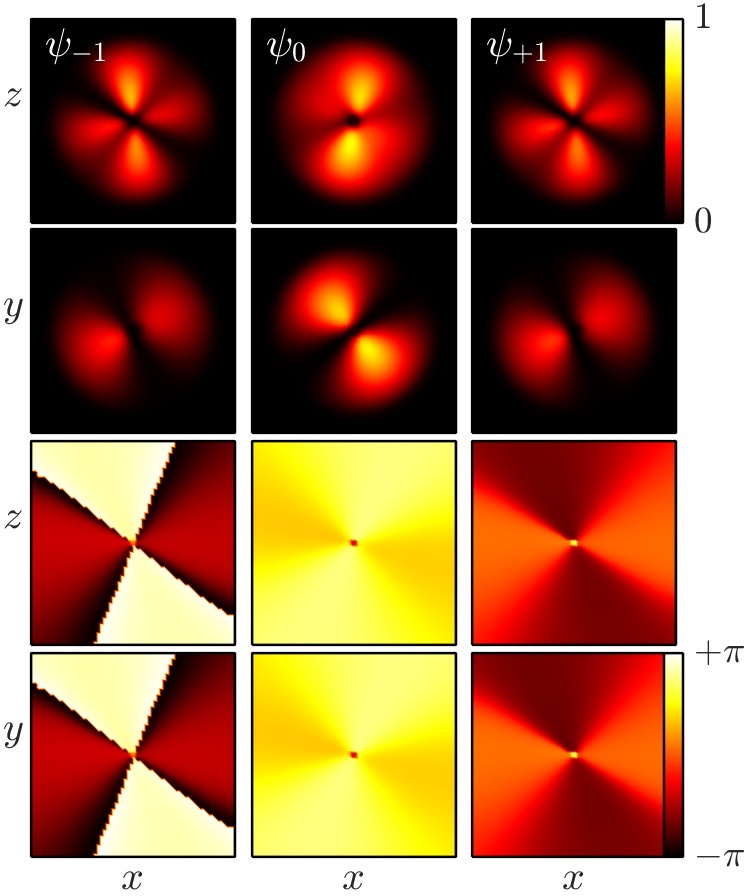}
}
\caption{
\label{fig:monopole_mu20_evals}
(Color online)
Most unstable eigenfunction for the monopole solution for $\mu=20$
Top row of panels:
isocontours for the density of most unstable eigenvector.
Bottom four rows:
cuts of the most unstable eigenvector. The first two rows correspond to density
cuts
and the last two to the corresponding phase.
All cuts pass through the origin.
Please refer to Appendices~\ref{app:BdG} and~\ref{app:numB} for details
on the BdG formulation and the numerical methods involved.
}
\end{figure}
%%%%%%%%%%%%%%%%%%%%%%%%%%%%%%%%%%%%%%%%%%%%%%%%%%%%%%%%%%%%%%%

%%%%%%%%%%%%%%%%%%%%%%%%%%%%%%%%%%%%%%%%%%%%%%%%%%%%%%%%%%%%%%%
\subsection{Stability}
%%%%%%%%%%%%%%%%%%%%%%%%%%%%%%%%%%%%%%%%%%%%%%%%%%%%%%%%%%%%%%%

Let us now characterize also the stability of the family of monopole solutions as
a function of the chemical potential $\mu$. For that purpose we utilize the
Bogoliubov-de Gennes (BdG), linear stability equations for small perturbations
away from stationary fixed points; see Appendix~\ref{app:BdG} for
details of the
perturbation ansatz and the corresponding stability matrix.
The outcome of this analysis yields the eigenvalues and eigenvectors
of the linearization around the equilibrium state (here, the monopole)
and showcases the potential spectral stability or instability of the
examined nonlinear state.

It is interesting to highlight that the bifurcation of
 the monopole from the linear limit eigenvalue of $\mu=5/2$ already
provides information about the number of potential instabilities of
this excited state in the way of so-called negative energy
modes~\cite{frantzeskakis2015defocusing}. Indeed, near the linear
limit, extending the eigenvalue calculation of the single component
case of Ref.~\cite{PhysRevA.62.053606}, we find that the BdG linearization
operator of Appendix~\ref{app:BdG} becomes a diagonal one with $\hat{h}_0-\mu$ 
and $-(\hat{h}_0-\mu)$ along the diagonals. In this case, we can directly
calculate the spectrum from the knowledge of the quantum harmonic oscillator
one as: $\pm (p + l + k-1)$ for $\mu=5/2$ (with both sets of
energies/eigenvalues measured in units of $\Omega$).

It is then straightforward to observe that the negative energy
modes in each component will arise due to the $(p,l,k)=(0,0,0)$ state and
hence, given the component multiplicity of components, there will be 3
such states. Similarly the 3 combinations $(1,0,0)$, $(0,1,0)$ and $(0,0,1)$
will result in vanishing eigenvalues for a total of 9 such, given the component
multiplicity. Finally, e.g., the positive eigenvalues pertaining to
$p+l+k=2$ will, through a similar count, be 6 per component and 18 in total.
The important count among these different ones is that of the negative
energy modes. Given the topological nature of such states, these modes
will preserve the sign of their energy {\it unless} they collide with positive
energy ones leading to an instability via the formation of a complex eigenvalue
quartet. Indeed, this scenario does materialize near the linear limit
for the monopole, where we find it to be unstable due to 3 complex
eigenvalue quartets (data not shown). Nevertheless, the more general conclusion
of relevance to all values of, e.g., the chemical potential is that such a
monopole state carries the potential for up to 3 instabilities via such
quartets at any value of $\mu$, due to the presence of these 3
negative energy modes.

Additionally, our computations show that the spectrum for the monopole
features an instability with a positive
real eigenvalue corresponding to an exponential instability.
A typical most unstable eigenfunction for the
monopole obtained on a $[-12,12]^{3}$ grid
with $51^3$ points is shown in Fig.~\ref{fig:monopole_mu20_evals}.
Closer examination of this eigenvector reveals that it
contains same signed lobes to either side of the original vortex lines of the
steady-state solution. These lobes, when added to the steady state, will
result in an effective displacement, along the $(x,y)$ plane, for the vortex
lines in the $\psi_{\pm1}$ components.
This displacement is typical when adding an even eigenfunction to an odd
solution (or vice-versa). It is important to note that, since the vortex lines
in the $\psi_{-1}$ and $\psi_{+1}$ components have opposite charge, the
displacement for each vortex line is in the opposite direction.
Therefore, we expect that the initial destabilization of the monopole
to be attributed to a symmetry breaking scenario where the vortex lines in the
$\psi_{\pm 1}$ components separate and drift away from the trap's center
(which will be corroborated through direct numerical integration; see below).

%%%%%%%%%%%%%%%%%%%%%%%%%%%%%%%%%%%%%%%%%%%%%%%%%%%%%%%%%%%%%%%
\begin{figure}[!htbp]
 \includegraphics[width=0.99\columnwidth]{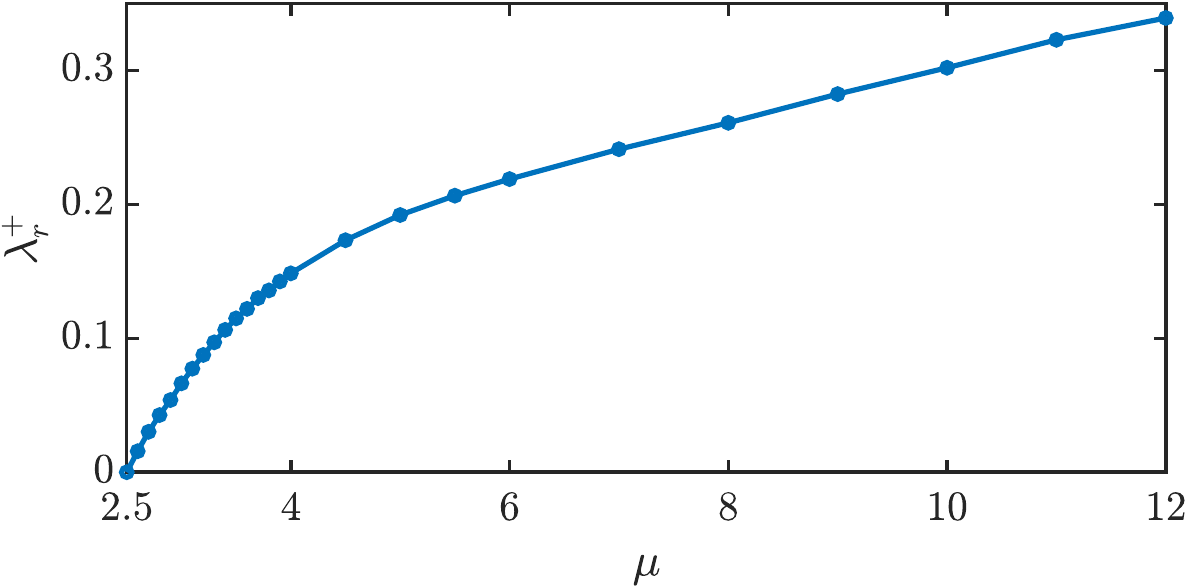}
\caption{
\label{fig:monopole_real_eval}
(Color online)
Most unstable (real) eigenvalue  for the monopole solution as
a function of the chemical potential $\mu$, depicting the growth
rate associated with the corresponding eigendirection.
}
\end{figure}
%%%%%%%%%%%%%%%%%%%%%%%%%%%%%%%%%%%%%%%%%%%%%%%%%%%%%%%%%%%%%%%

Figure~\ref{fig:monopole_real_eval} depicts the instability growth rate,
through the positive real part, denoted as $\lambda_r^+$ hereafter of the most
unstable eigenvalue, of the monopole solution as a function of the chemical
potential (see Appendix \ref{app:numB} for details on the numerical computations).
As the figure indicates, as one gets closer to the linear
limit where the monopole solution is born ($\mu=5/2$), the real part of
the relevant eigenvalue tends to zero.
Besides the unstable real eigenvalue (and its negative
sibling owing to the Hamiltonian nature of the problem)
with triple multiplicity,
the spectrum is composed of a set of purely imaginary eigenvalues and
a zero eigenvalue with multiplicity eight.
The zero eigenvalues are associated with the symmetries (or invariances)
of the corresponding steady-state solution and equations. 
In particular, we note in
that connection that the monopole can be rotated around any direction,
in addition to its possessing an overall phase invariance,
associated with the total atom number conservation law.
On the other hand, the purely imaginary eigenvalues consist of
(i) the spectrum related to the ground state of the system and
(ii) negative energy modes associated with the excited state nature
of the monopole configuration of interest here.

%%%%%%%%%%%%%%%%%%%%%%%%%%%%%%%%%%%%%%%%%%%%%%%%%%%%%%%%%%%%%%%
\begin{figure}[!htbp]
\includegraphics[height=4.3cm]{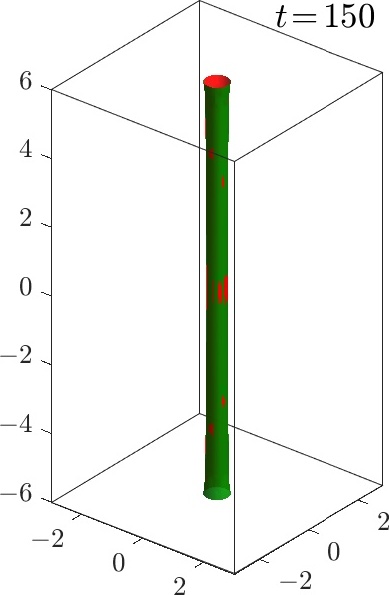}
\includegraphics[height=4.3cm]{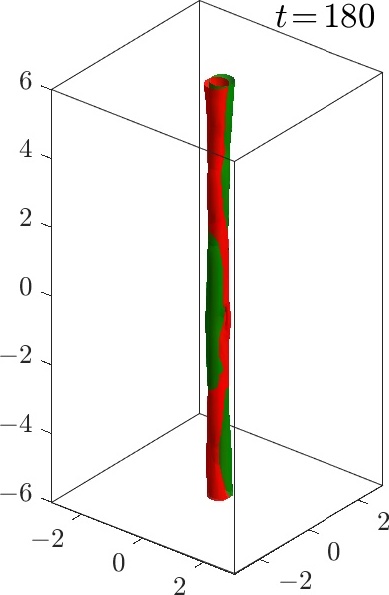}
\includegraphics[height=4.3cm]{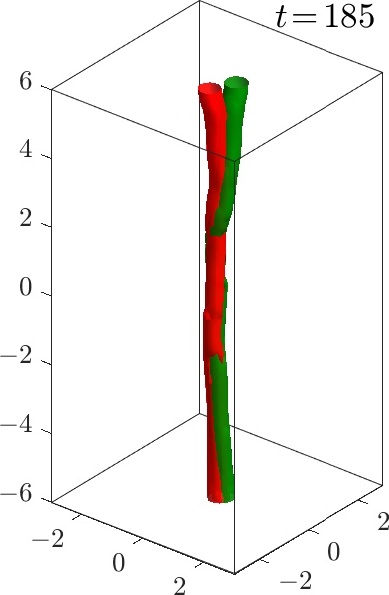}
\includegraphics[height=4.3cm]{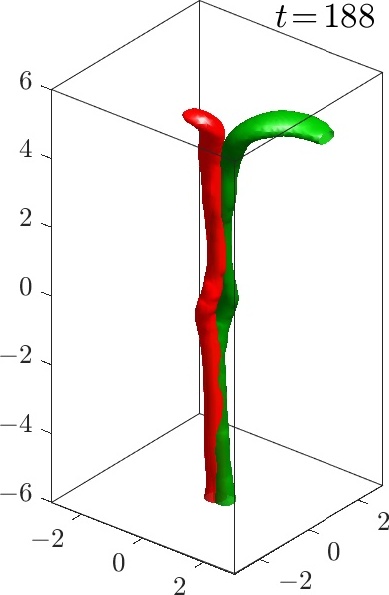}
\includegraphics[height=4.3cm]{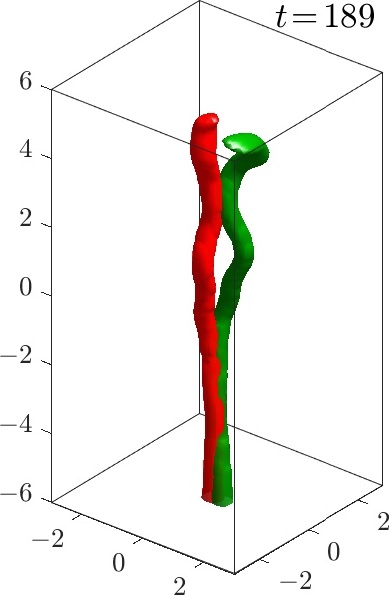}
\includegraphics[height=4.3cm]{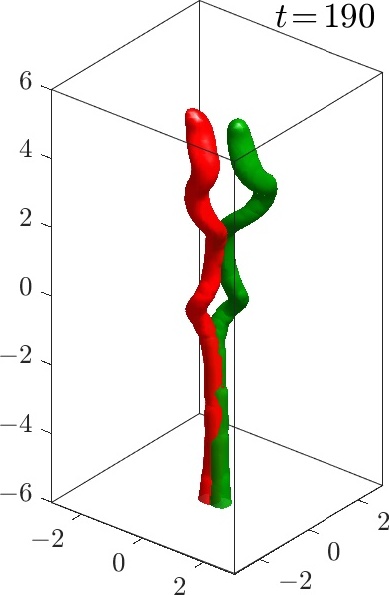}
\caption{
\label{fig:monopole_mu20_dyn}
(Color online)
Destabilization dynamics for the monopole solution with $\mu=20$ at
the indicated times. Shown are overlaid isocontours for the vorticity
corresponding to the $\psi_{+1}$ (red) and $\psi_{-1}$ (green) components.
The system was initialized with the numerically exact steady state monopole
solution with a spatial random perturbation of size $10^{-8}$
(relative to the steady state size).
For the corresponding movie please follow this
\href{https://drive.google.com/file/d/19ulkw52LKY6VJLNWVi3VE9L-5oaJpDDE}{link}.
}
\end{figure}
%%%%%%%%%%%%%%%%%%%%%%%%%%%%%%%%%%%%%%%%%%%%%%%%%%%%%%%%%%%%%%%

%%%%%%%%%%%%%%%%%%%%%%%%%%%%%%%%%%%%%%%%%%%%%%%%%%%%%%%%%%%%%%%
\begin{figure}[!htbp]
\includegraphics[height=4.3cm]{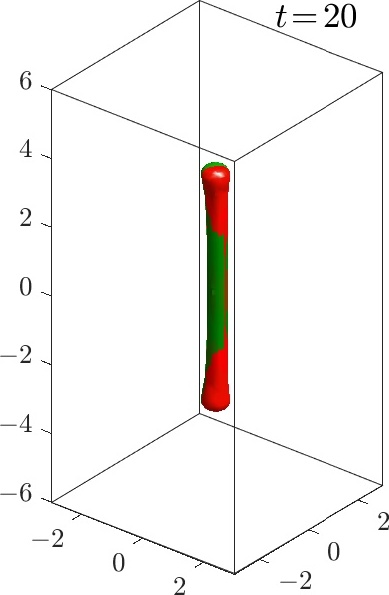}
\includegraphics[height=4.3cm]{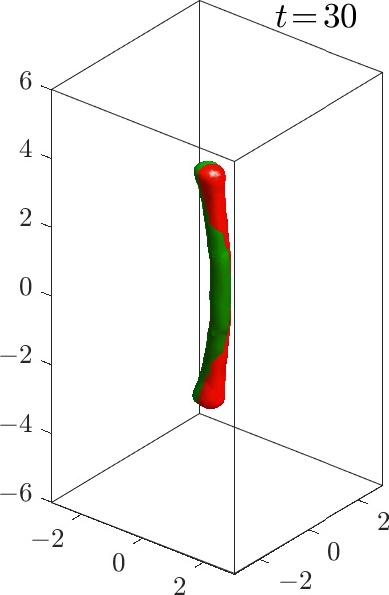}
\includegraphics[height=4.3cm]{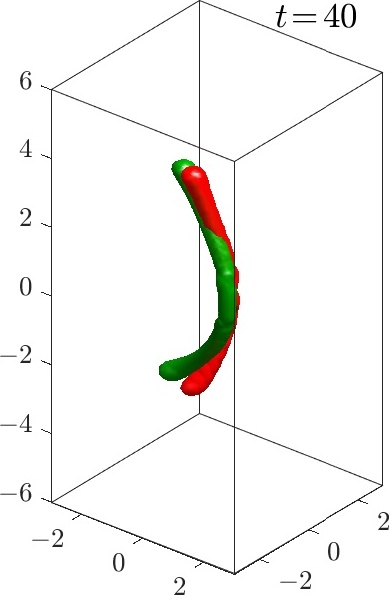}
\includegraphics[height=4.3cm]{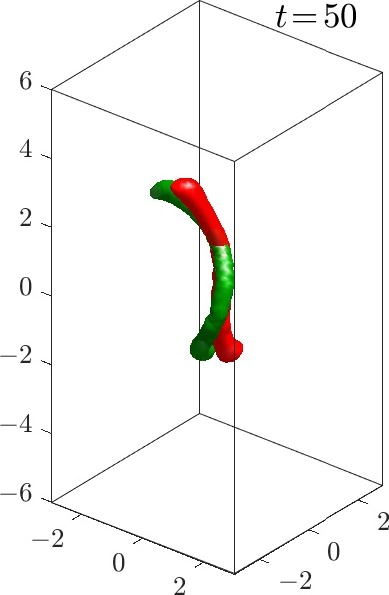}
\includegraphics[height=4.3cm]{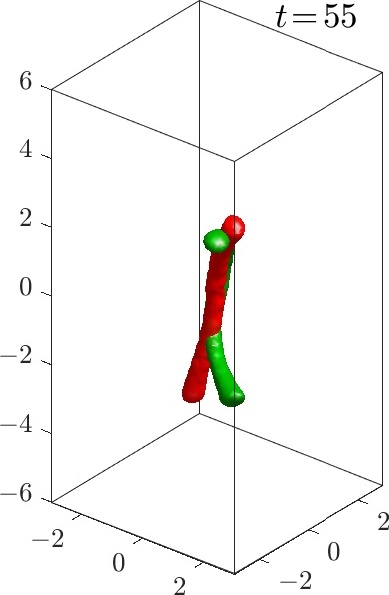}
\includegraphics[height=4.3cm]{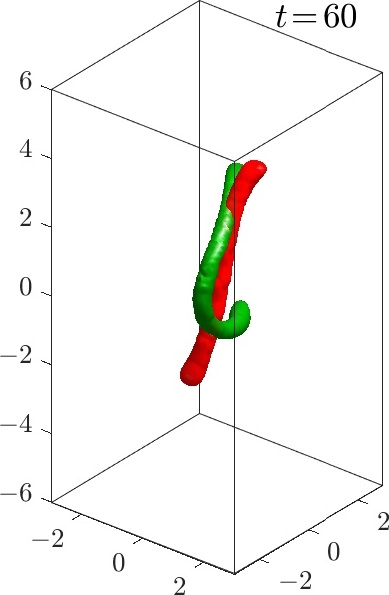}
\caption{
\label{fig:monopole_mu04_dyn}
(Color online)
Same as in Fig.~\ref{fig:monopole_mu20_dyn} but for $\mu=4$.
For the corresponding movie please follow this
\href{https://drive.google.com/file/d/1Jy3Lh6DmFc-KXmUc_Zsk3l3oQmpxNL71}{link}.
}
\end{figure}
%%%%%%%%%%%%%%%%%%%%%%%%%%%%%%%%%%%%%%%%%%%%%%%%%%%%%%%%%%%%%%%

%%%%%%%%%%%%%%%%%%%%%%%%%%%%%%%%%%%%%%%%%%%%%%%%%%%%%%%%%%%%%%%
\subsection{Dynamics}
%%%%%%%%%%%%%%%%%%%%%%%%%%%%%%%%%%%%%%%%%%%%%%%%%%%%%%%%%%%%%%%

Let us now follow the dynamical destabilization of the monopole solution by
directly integrating the equations of motion initialized by
slightly (randomly) perturbed stationary configurations.
The cases for $\mu=20$ and $\mu=4$ are depicted in Figs.~\ref{fig:monopole_mu20_dyn}
and~\ref{fig:monopole_mu04_dyn}, respectively. As we are mainly
interested in the fate of the vortex lines present in the $\psi_{\pm1}$
components, we only show overlays of their corresponding isocontours of vorticity.
In line with the eigenvector stability results above, we see that the
destabilization of the monopole evolves through a symmetry breaking between
the vortex lines in the $\psi_{\pm1}$ components.
We typically observe that the vortex lines slightly split from one
another and start slowly drifting away from the $z$-axis. 
The drift is tantamount (albeit not exactly equivalent, since here the 
vortex lines pertain to different components) to the behavior of adjacent 
vortex lines in one-component BECs. There, vortex lines of same charge 
tend to rotate around each other while vortex lines of opposite charge 
(as is the case for our system) tend to travel parallel to each other.
In general we observed that, eventually, the vortex
lines separate further and more complex dynamics ensues, particularly
for higher values of $\mu$ where the vortex lines are thin and
prone to undulations (Kelvin waves)~\cite{frantzeskakis2015defocusing}.
However, we would like to focus on a particularly prevalent feature that is
common to most monopole configuration destabilization dynamics for high
enough chemical potential.

As an example, close inspection of the destabilization
dynamics for $\mu=20$ (see Fig.~\ref{fig:monopole_mu20_dyn}) reveals that
the vortex lines tend to mainly split from each other at the edges of the
cloud and, more importantly, around the center of the trap. The split at the
cloud's periphery is eventually responsible for the ``peeling'' off of the
two vortex lines and the eventual complex vortex line dynamics.
However, before entering this fully developed
filamentary dynamics, the split between the vortex lines at the
center of the trap creates a small bulge in the shape of a ring composed
of two halves, one from each of the $\psi_{\pm 1}$ components.
Namely, as clearly seen in the $t=188$ panel of Fig.~\ref{fig:monopole_mu20_dyn},
the vortex lines develop a ring around the trap's center. This
configuration is reminiscent of an AR, first studied
in detail in the present BEC context in Ref.~\cite{Ruostekoski_2003},
which we study in more detail in the next section of this work.

It is interesting that as part of the destabilization of the monopole,
due to the central splitting of the vortex lines across the $\psi_{\pm 1}$
components, the dynamics tend to ``hover'' momentarily through a
solution akin to an AR.
While transiently we observe such a ring
configuration in a segment of the condensate, over long time scales
the dynamics of the monopole involves multiple recurrences and
an eventual split of the vortex lines and a disappearance of the
associated vorticity towards the background (TF) cloud.
It is also relevant to comment that the
dynamics of Fig.~\ref{fig:monopole_mu04_dyn} similarly illustrate
the peeling off of the vortex lines at the edges, yet are far less suggestive
towards the creation of a potential ring near the center of the trap.
We attribute this to the important feature (as we will see below) stemming
from our analysis of ARs that such structures do {\em not}
exist at such low values of the chemical potential. Hence, they are
no longer natural candidates for such transient dynamical observations
in this case of lower chemical potential.

%%%%%%%%%%%%%%%%%%%%%%%%%%%%%%%%%%%%%%%%%%%%%%%%%%%%%%%%%%%%%%%
\begin{figure}[!htbp]
 \includegraphics[width=0.99\columnwidth]{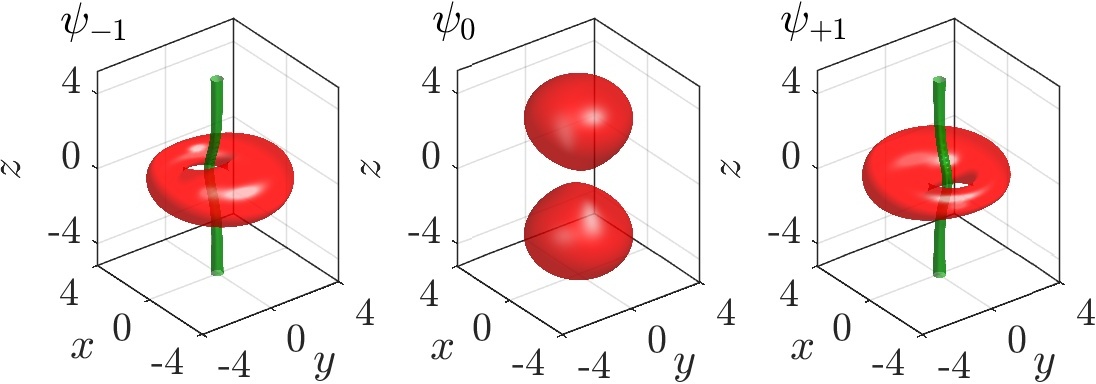}
\\[3.0ex]
 \includegraphics[width=0.80\columnwidth]{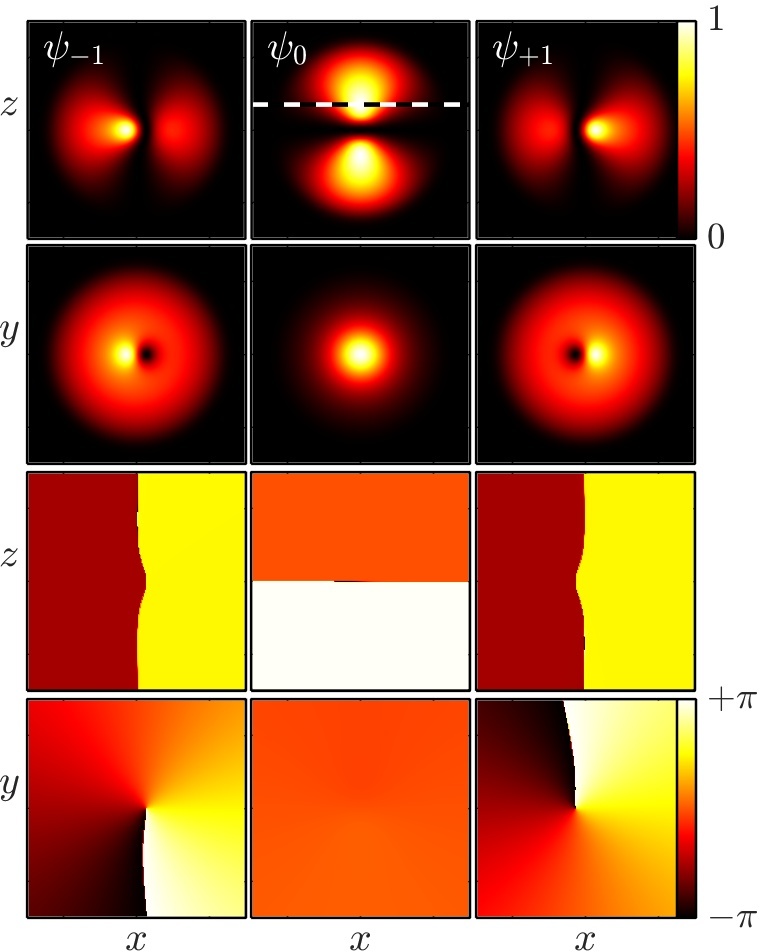}
 \caption{
\label{fig:alice_mu20}
(Color online)
First Alice ring (AR) solution for $\mu=20$.
Same layout as in Fig.~\ref{fig:monopole_mu20}, with isolevels of
density (red) and vorticity (green) in the top row and different cuts,
clearly showcasing the ``symmetry breaking'' nature of the AR
structure are shown in the 2nd-5th rows along the different planes and
for the different components.}
\end{figure}
%%%%%%%%%%%%%%%%%%%%%%%%%%%%%%%%%%%%%%%%%%%%%%%%%%%%%%%%%%%%%%%

%%%%%%%%%%%%%%%%%%%%%%%%%%%%%%%%%%%%%%%%%%%%%%%%%%%%%%%%%%%%%%%
\begin{figure}[!htbp]
 \includegraphics[width=0.99\columnwidth]{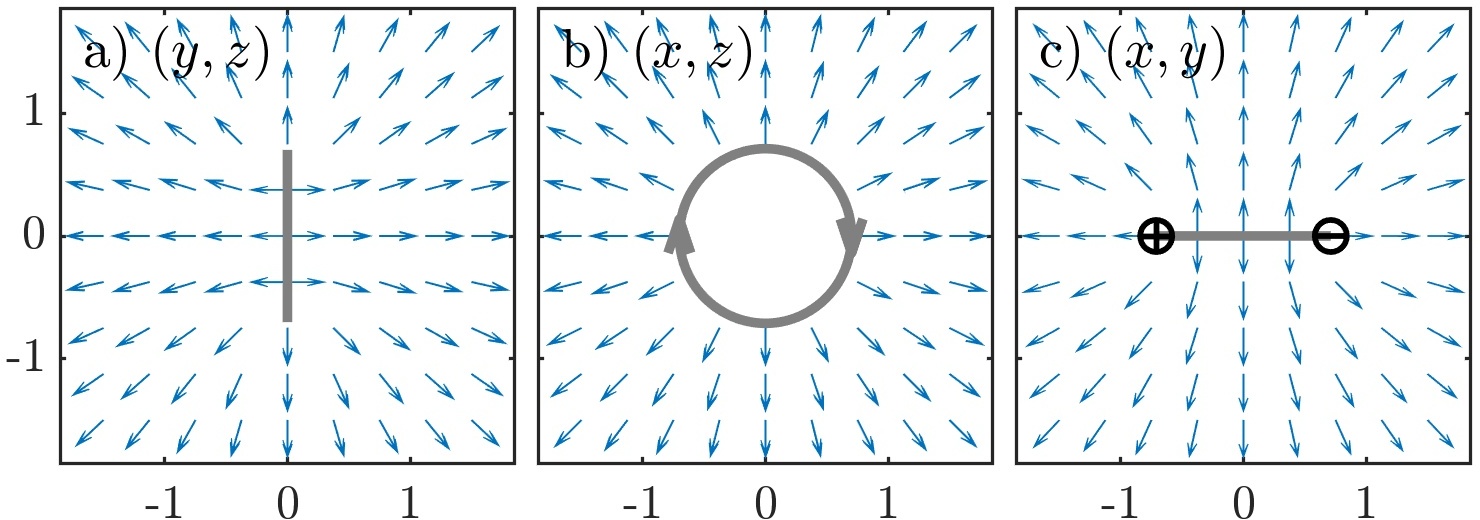}
\caption{
\label{fig:alice_nematic}
Nematic vector for the first AR solution for $\mu=20$.
In contrast with the monopole depicted in Fig.~\ref{fig:monopole_nematic},
the AR has a disk (inside the ring itself) where the nematic vector field
abruptly reverses direction. 
Nonetheless, the AR's far field does match the one for the monopole.
The grey curves depict the location of the AR and the `$+$' and
`$-$' symbols in panel c) highlight the location of the vortices in
the $\psi_+$ and $\psi_-$ components, respectively.
}
\end{figure}
%%%%%%%%%%%%%%%%%%%%%%%%%%%%%%%%%%%%%%%%%%%%%%%%%%%%%%%%%%%%%%%

%%%%%%%%%%%%%%%%%%%%%%%%%%%%%%%%%%%%%%%%%%%%%%%%%%%%%%%%%%%%%%%
\begin{figure}[!htbp]
 \includegraphics[width=0.99\columnwidth]{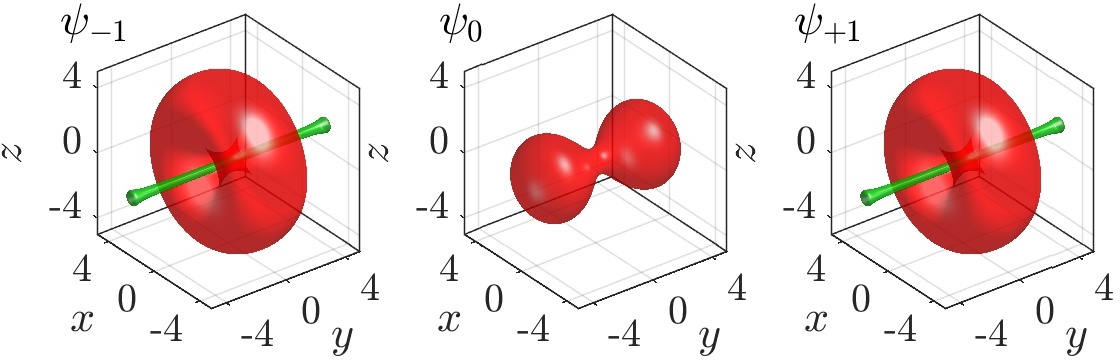}
\\[3.0ex]
 \includegraphics[width=0.80\columnwidth]{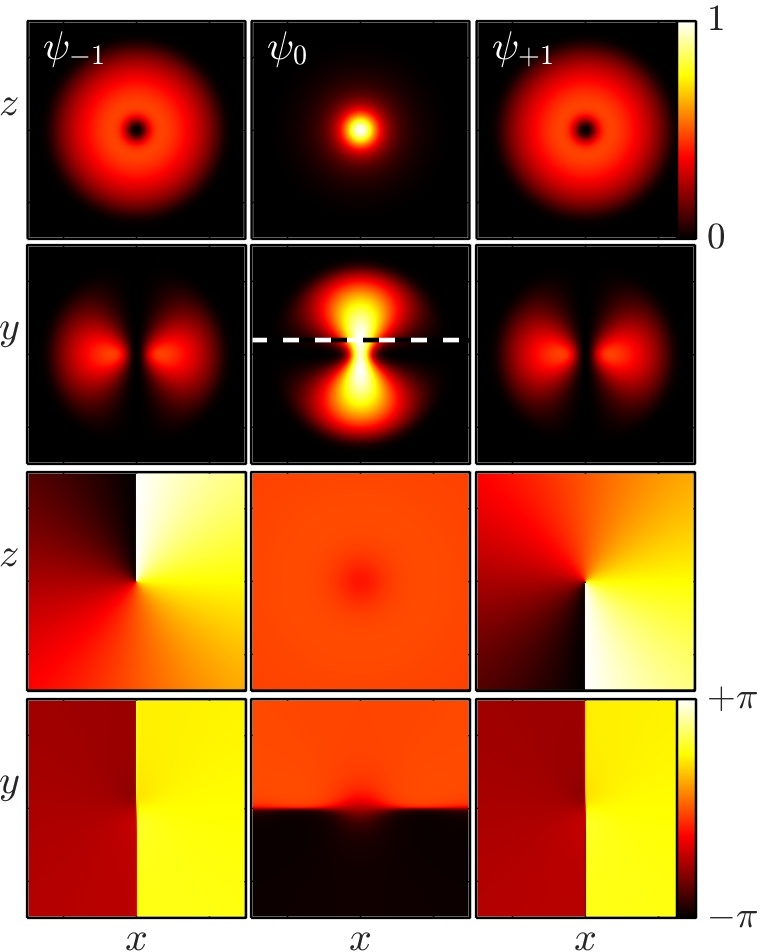}
\\[3.0ex]
\includegraphics[width=0.83\columnwidth]{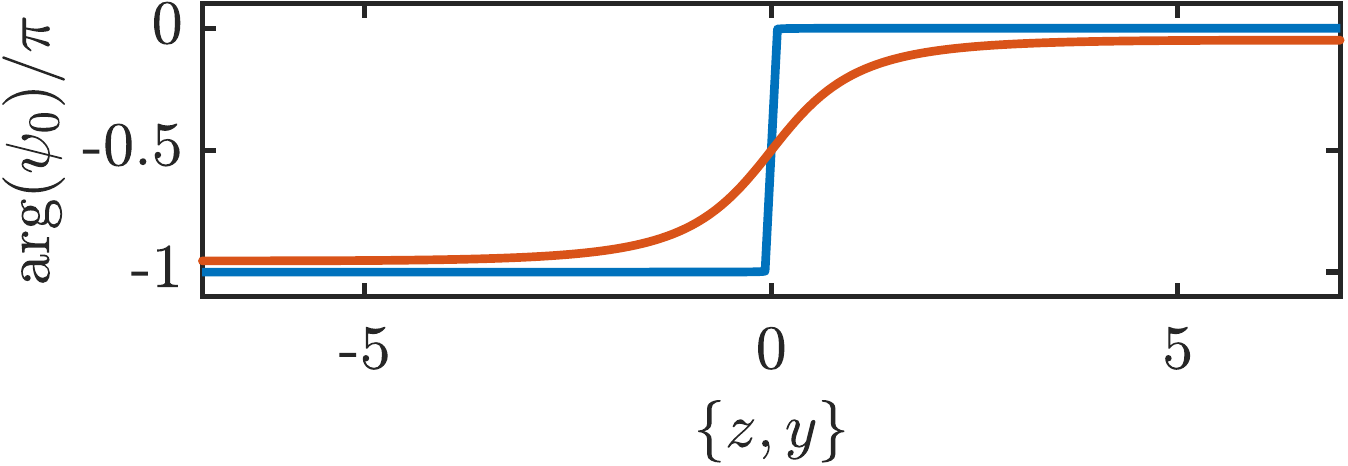}
\qquad\qquad\quad
 \caption{
\label{fig:alice_ROT_mu20}
(Color online)
First Alice ring solution for $\mu=20$ with quantization axis perpendicular 
to the plane of the ring (i.e., parallel to the $y$-axis). In this basis it 
is apparent that the superfluid density is nonzero and the phase changes 
smoothly by approximately $\pi$ in the region along the $y$-axis. 
This solution is obtained by applying a rotation in spin space
with Euler angles $\alpha=\gamma=0$ and $\beta=\pi/2$
(cf.~Eq.~(551) in Ref.~\cite{KAWAGUCHI2012253}).
Top panel: same layout as in Fig.~\ref{fig:alice_mu20}.
Bottom panel: phase corresponding to $\psi_0(0,0,z)$ (blue) and 
$\psi_0(0,y,0)$ (orange) for the first AR in, respectively, spinor
coordinates (cf.~Fig.~\ref{fig:alice_mu20}) and with the
quantization axis parallel to the $y$-axis (see top panels).
}
\end{figure}
%%%%%%%%%%%%%%%%%%%%%%%%%%%%%%%%%%%%%%%%%%%%%%%%%%%%%%%%%%%%%%%

%%%%%%%%%%%%%%%%%%%%%%%%%%%%%%%%%%%%%%%%%%%%%%%%%%%%%%%%%%%%%%%
\begin{figure*}[!htbp]
\includegraphics[width=\textwidth]{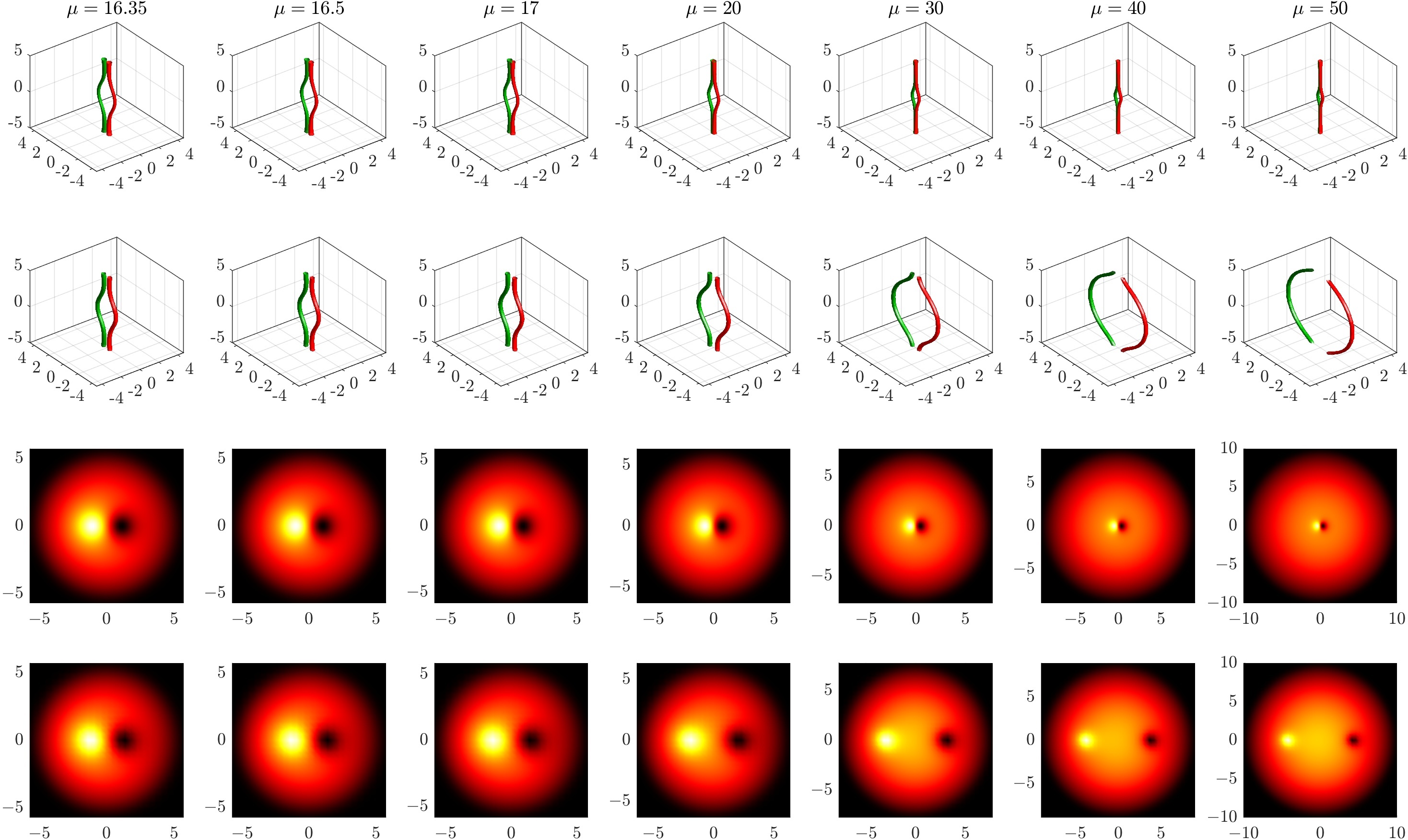}
\caption{
\label{fig:alice_mus}
(Color online)
AR steady states for different values of $\mu$ as indicated in the panels.
The first and third row of panels correspond to the first
AR branch (smaller rings) while the second and fourth rows correspond to
the second AR branch (larger rings).
The top two rows of panels depict the isolevel cuts of constant
vorticity for the $\psi_{+1}$ (green) and $\psi_{-1}$ (red) components.
The bottom two rows of panels depict the corresponding $z=0$ cut of the density
for the $\psi_{-1}$ component.
Note that the plotting window for the density cuts
is rescaled by the TF radius $R_{\rm TF}$ as
$[-1.5R_{\rm TF},1.5R_{\rm TF}]\thintimes [-1.5R_{\rm TF},1.5R_{\rm TF}]$
to keep the apparent size of the cloud constant.
}
\end{figure*}
%%%%%%%%%%%%%%%%%%%%%%%%%%%%%%%%%%%%%%%%%%%%%%%%%%%%%%%%%%%%%%%

%%%%%%%%%%%%%%%%%%%%%%%%%%%%%%%%%%%%%%%%%%%%%%%%%%%%%%%%%%%%%%%
\section{Alice Rings}
\label{sec:ARs}
%%%%%%%%%%%%%%%%%%%%%%%%%%%%%%%%%%%%%%%%%%%%%%%%%%%%%%%%%%%%%%%

An AR corresponds to a half-quantum vortex ring which resembles
a monopole solution~\cite{KAWAGUCHI2012253,Ruostekoski_2003} far
from the origin  of the structure.  In this structure,
a $\pi$ change in the global phase $\phi$ is accompanied by a
$\pi$-disclination of the quantization vector $\hat{\mathbf{d}}$.
Indeed, the singular point defect of the 't Hooft-Polyakov monopole
setting deforms itself into a ring structure where half the ring
consists of the (deformed) vortex line of the $\psi_1$ component, while the
other half of the (symmetrically deformed in the opposite
direction) vortex line of the $\psi_{-1}$ component.
Thus, it is not surprising to expect that, as already hinted above,
that the dynamical instability of the monopole is intimately connected
with the dynamical emergence of the AR.

%%%%%%%%%%%%%%%%%%%%%%%%%%%%%%%%%%%%%%%%%%%%%%%%%%%%%%%%%%%%%%%
\subsection{Steady States}
%%%%%%%%%%%%%%%%%%%%%%%%%%%%%%%%%%%%%%%%%%%%%%%%%%%%%%%%%%%%%%%

Inspired by the appearance of the vortex line bulge close to the center in the
dynamical destabilization of the monopole solution, we initialize our steady-state
optimizer with a snapshot of the monopole destabilization when the
bulge is visible but before the vortex lines bend and twist
themselves, substantially detaching from each other.
By doing so, we are able to identify a {\it stationary} AR solution and follow it for
different values of the chemical potential.
For instance, Fig.~\ref{fig:alice_mu20} depicts the AR solution
for $\mu=20$ where we clearly see the bulge of the vortex lines in
opposite directions for the $\psi_{+1}$ and $\psi_{-1}$ components.
We call this solution the {\em first} AR since, as we will reveal below,
there is another branch of AR solutions (the {\em second} AR).
Figure~\ref{fig:alice_nematic} depicts the nematic vector for a typical AR.
The nematic vector field shows that, as the vortex lines in the $\psi_{\pm1}$
open up and create a ring, the far field still corresponds to the hedgehog
(radially outward) structure of a monopole. The principal new feature in this
representation is the presence of a
$\pi$-{\em discontinuity} of the nematic vector across the plane of the ring.
Although the nematic (director) vector is discontinuous on the AR disk, the $\mathbb{Z}_2$ 
nematic symmetry $(\hat{\mathbf{d}},\phi) \rightarrow (-\hat{\mathbf{d}},\phi+\pi)$ 
permits $\Psi$ to remain continuous provided the discontinuity in the direction 
of $\hat{\mathbf{d}}$ is accompanied by a discontinuity of $\pi$ in the phase $\phi$.
Given that the director orientation and the phase are thus not uniquely prescribed, 
we return to the full solution of $\Psi$ for an in-depth account of the AR.

The crucial topology of the AR manifests itself in the nonzero superfluid density 
along the $y$-axis, and perpendicular to the plane of the ring, which disrupts the 
solitonic structure associated with its parent monopole. This feature is most easily 
discerned in the $m=0$ spinor component density in a basis quantized along the 
$y$-axis (see Fig.~\ref{fig:alice_ROT_mu20}). Moreover, the condensate phase 
changes continuously by approximately $\pi$ along the axis 
(see orange curve in bottom panel), as is expected for a half-quantum vortex ring
---contrast this phase variation with that for the AR represented in the original 
spinor basis of Fig.~\ref{fig:alice_mu20} which displays a discontinuous jump over 
its axis (see blue curve in the bottom panel of Fig.~\ref{fig:alice_ROT_mu20}). 
It is interesting to note how the two different 
choices of quantization axis reveal complementary information: when quantized along $y$ 
(Fig.~\ref{fig:alice_ROT_mu20}) the nonzero density in the $m=0$ spinor 
component defines the radius of the ring in the $(x,z)$ plane and the $\pi$ 
phase winding, whereas when quantized along $z$ (Fig.~\ref{fig:alice_mu20}) 
the $m=\pm 1$ spinor components in the $(x,y)$ plane feature a structure 
reminiscent of a half-quantum vortex dipole.
Also, interestingly, in the former case, the vortex
lines in the $m=\pm 1$ components end up aligned, while in the latter
feature a misaligned portion where each component traces
half of the AR. These complementarities are reminiscent
of the vortex-vortex states and their SU(2) rotations considered
earlier in Ref.~\cite{stathis}.
All of the obtained AR solutions for the different
parameters considered (variations of the chemical potential and trapping strengths), 
including the second AR branch (see below), display this general structure.

%%%%%%%%%%%%%%%%%%%%%%%%%%%%%%%%%%%%%%%%%%%%%%%%%%%%%%%%%%%%%%%
\begin{figure}[!htbp]
 \includegraphics[width=0.99\columnwidth]{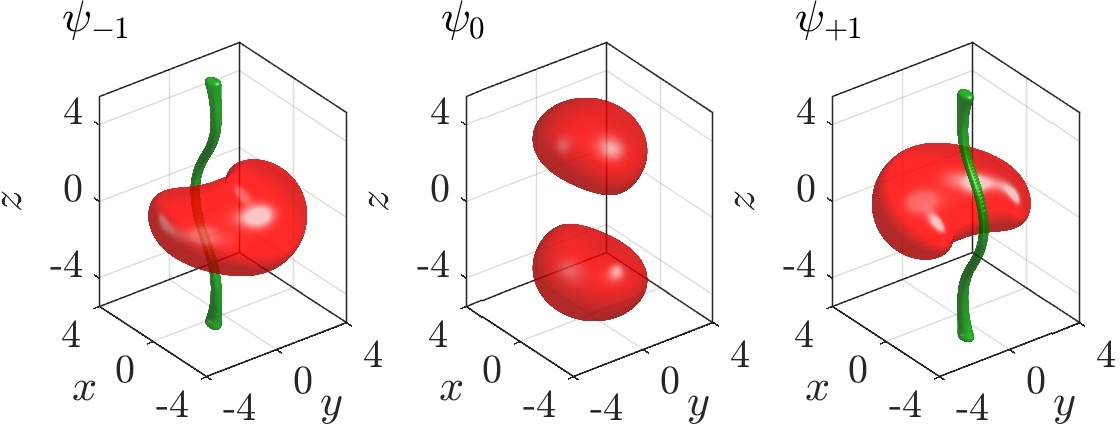}
\\[3.0ex]
 \includegraphics[width=0.80\columnwidth]{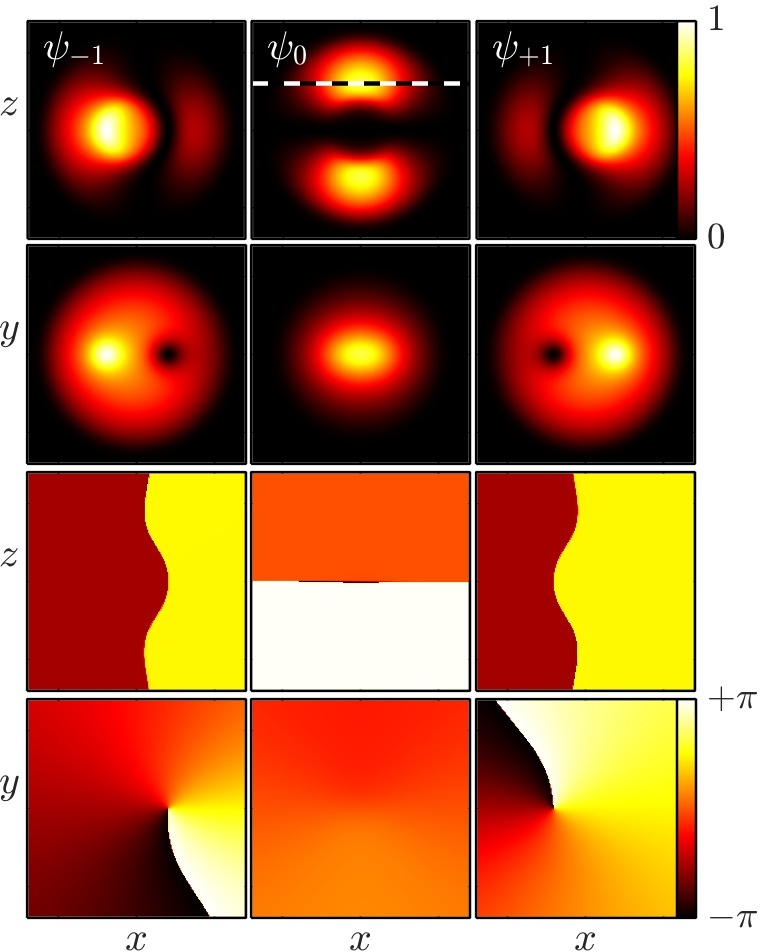}
 \caption{
\label{fig:alice_2nd_mu20}
(Color online)
Second AR solution for $\mu=20$.
Same layout as in Fig.~\ref{fig:alice_mu20}.
}
\end{figure}
%%%%%%%%%%%%%%%%%%%%%%%%%%%%%%%%%%%%%%%%%%%%%%%%%%%%%%%%%%%%%%%

%%%%%%%%%%%%%%%%%%%%%%%%%%%%%%%%%%%%%%%%%%%%%%%%%%%%%%%%%%%%%%%
\begin{figure}[!htbp]
 \includegraphics[width=0.99\columnwidth]{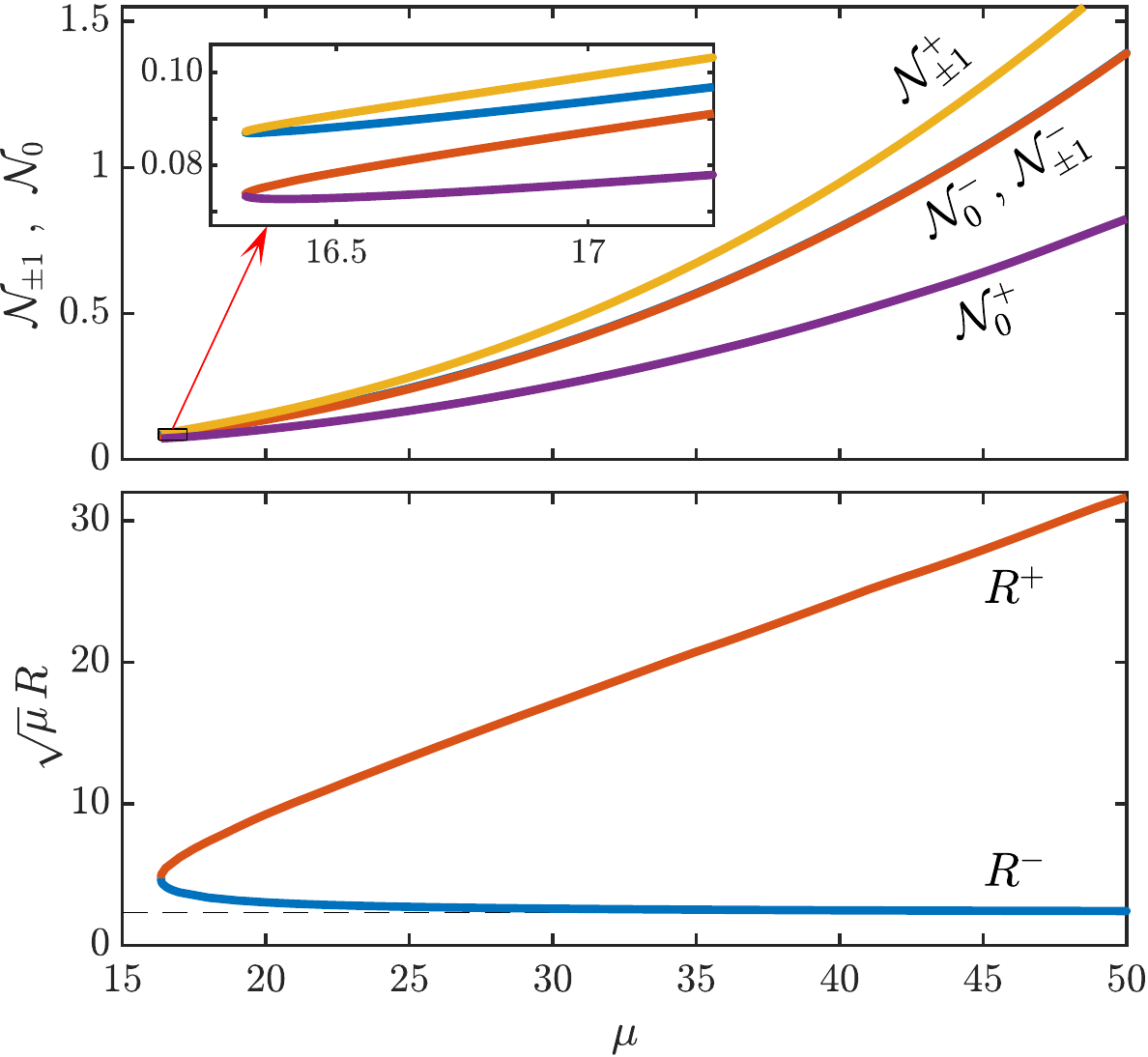}
\caption{
\label{fig:alice_mass_radius}
(Color online)
Top:
(adimensionalized) Atom numbers (\ref{eq:masseach}) for the AR
solution as the (adimensionalized) chemical potential $\mu$ is varied.
The AR solution seems to terminate, as $\mu$ decreases, at $\mu\approx 16.35$.
Bottom:
AR's radius vs.~chemical potential $\mu$.
The radius has been normalized by $1/\sqrt{\mu}$.
The horizontal  dashed line corresponds to $R=2.3/\sqrt{\mu}$.
The $-$ and $+$ superscripts correspond, respectively, to the
first (small) and second (large) AR solutions.
}
\end{figure}
%%%%%%%%%%%%%%%%%%%%%%%%%%%%%%%%%%%%%%%%%%%%%%%%%%%%%%%%%%%%%%%

Let us now discuss our parametric variation of the relevant AR waveforms.
Once we obtained a genuine AR solution
for a particular value of the chemical potential $\mu$, we used numerical
continuation to follow this (first) solution branch for different values
of $\mu$ as depicted in the first and third rows in Fig.~\ref{fig:alice_mus}.
Interestingly, the first AR solution branch seems to not exist for
values of $\mu<16.35$ and, thus, cannot be constructed all the way down
to the linear (small mass) limit. Therefore, we sought a companion branch
of solutions using arclength continuation~\cite{doedel_book}
around $\mu<16.35$, upon identifying that this branch of solutions
spontaneously emerges (out of the ``blue sky'') for 
values of $\mu \geq  16.35$.

The arclength continuation reveals that indeed the first AR
solution branch is connected to a second AR branch through a
saddle-center bifurcation.
The two AR solution branches
collide at the relevant turning point.
After going over the fold using arclength continuation, we are
able to follow the second AR branch for larger values of $\mu$.
For instance, Fig.~\ref{fig:alice_2nd_mu20} depicts a sample configuration
of this second AR solution for $\mu=20$.
More elements of this second AR branch are depicted in the second
and fourth rows of Fig.~\ref{fig:alice_mus}.
It is interesting that the first branch corresponds
to smaller rings when compared to the second branch and that, as $\mu$ increases,
the smaller rings of the first branch get smaller while the larger rings of
the second branch get larger. I.e., it is interesting to observe that
in a sense the AR emerges at a ``critical radius'' for $\mu\approx 16.35$
and thereafter the first branch approaches progressively
(for increasing chemical potential) the monopole solution with the
ring ``closing in'' by shrinking its radius, while the second branch
progressively tends to grow with the curved
segments of the vortex lines expanding toward the periphery of the
cloud. However, it is relevant to
keep in mind that the TF cloud edge also expands as $\mu$ 
increases (see also the comment below).

We have confirmed that the same qualitative behavior for the nematic vector,
as depicted for the first AR in Fig.~\ref{fig:alice_nematic}, is also present for the
AR members of the second family (results not shown here).
In order to better depict the
two AR branches within the same ``bifurcation diagram'',
the top panel of Fig.~\ref{fig:alice_mass_radius}
depicts the (dimensionless) atom number of the AR branches as a
function of $\mu$.
The bottom panel of Fig.~\ref{fig:alice_mass_radius} depicts the radius of
the (smaller) first ($R^-$) and (larger) second ($R^+$) ARs
normalized by $1/\sqrt{\mu}$.
The ring radii are extracted by determining the position of the positively
and negatively charged vortices present in the 2D cuts of the $\psi_\pm$
components on the $z=0$ plane.
Apparently, the rescaled radius $R^+$ for the second AR family increases
linearly with $1/\sqrt{\mu}$ as $\mu$ increases.
This indicates that $R^+\propto \sqrt{\mu}$ which means that indeed
$R^+$ it increases linearly with the size
(Thomas-Fermi radius $R_{\rm TF}\propto \sqrt{\mu}$)
of the condensate cloud ---this is also apparent
in the larger $\mu$ yet $z=0$ cuts in the fourth column of panels in
Fig.~\ref{fig:alice_mus} where the relative position of the vortices with
respect to condensate cloud edge seems to be constant.
In contrast to the enlarging behavior of the second AR family,
the first ring family tends to shrink as $\mu$ increases. In fact, the first
AR radius seems to tend to a constant fraction of $1/\sqrt{\mu}$ as
$\mu\rightarrow+\infty$.

%%%%%%%%%%%%%%%%%%%%%%%%%%%%%%%%%%%%%%%%%%%%%%%%%%%%%%%%%%%%%%%
\begin{figure}[!htbp]
\center{
\hspace{-0.5cm}
 \includegraphics[width=0.99\columnwidth]{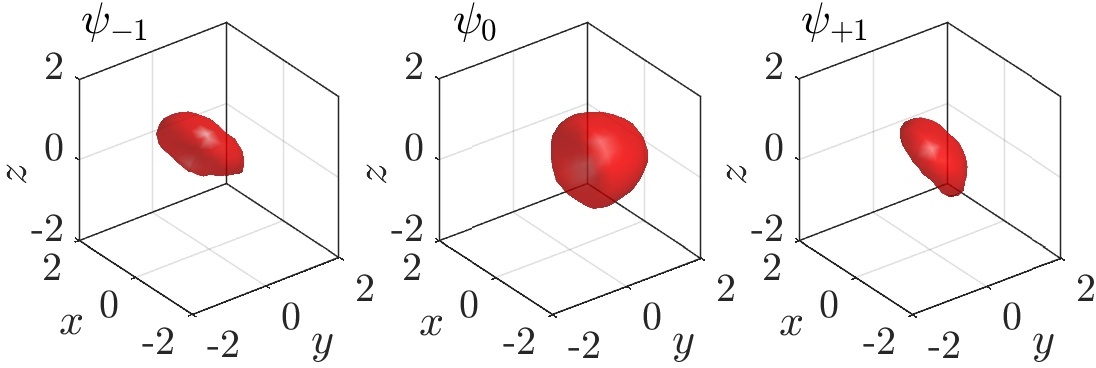}
\\[2.0ex]
\hspace{0.5cm}
 \includegraphics[width=0.80\columnwidth]{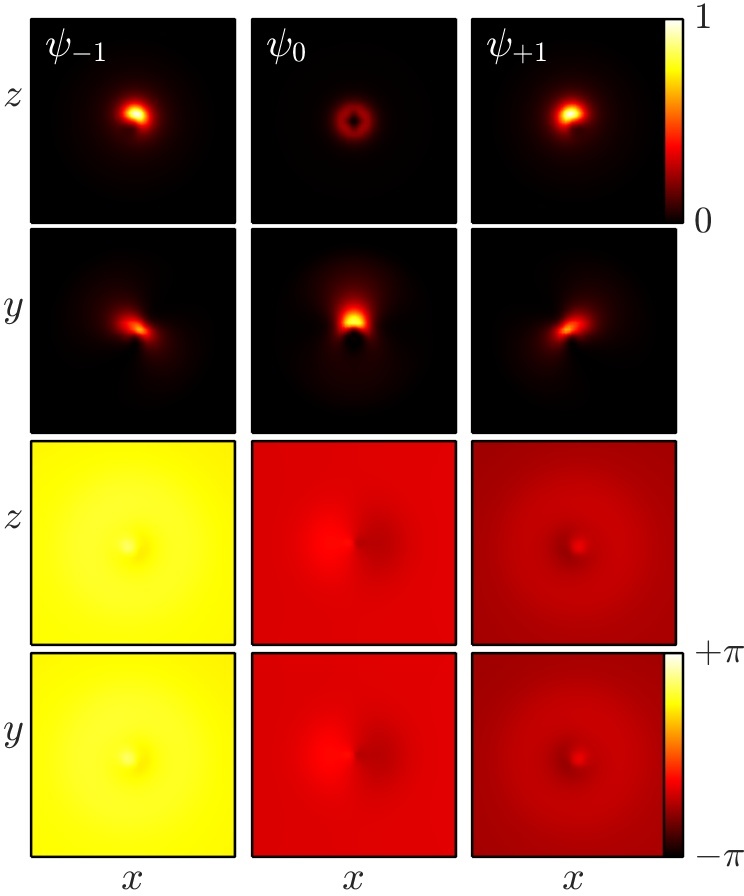}
}
\caption{
\label{fig:alice_mu20_evals}
(Color online)
Most unstable eigenfunction
for the first AR solution for $\mu=20$;
same layout as in Fig.~\ref{fig:monopole_mu20_evals}.
The most unstable eigenvector does not carry a
sizeable amount of vorticity and thus isocontours of
vorticity have been omitted.
Please refer to Appendices \ref{app:BdG} and \ref{app:numB} for details
on the BdG formulation and the numerical methods involved.
}
\end{figure}
%%%%%%%%%%%%%%%%%%%%%%%%%%%%%%%%%%%%%%%%%%%%%%%%%%%%%%%%%%%%%%%

%%%%%%%%%%%%%%%%%%%%%%%%%%%%%%%%%%%%%%%%%%%%%%%%%%%%%%%%%%%%%%%
\begin{figure}[!htbp]
 \includegraphics[width=0.99\columnwidth]{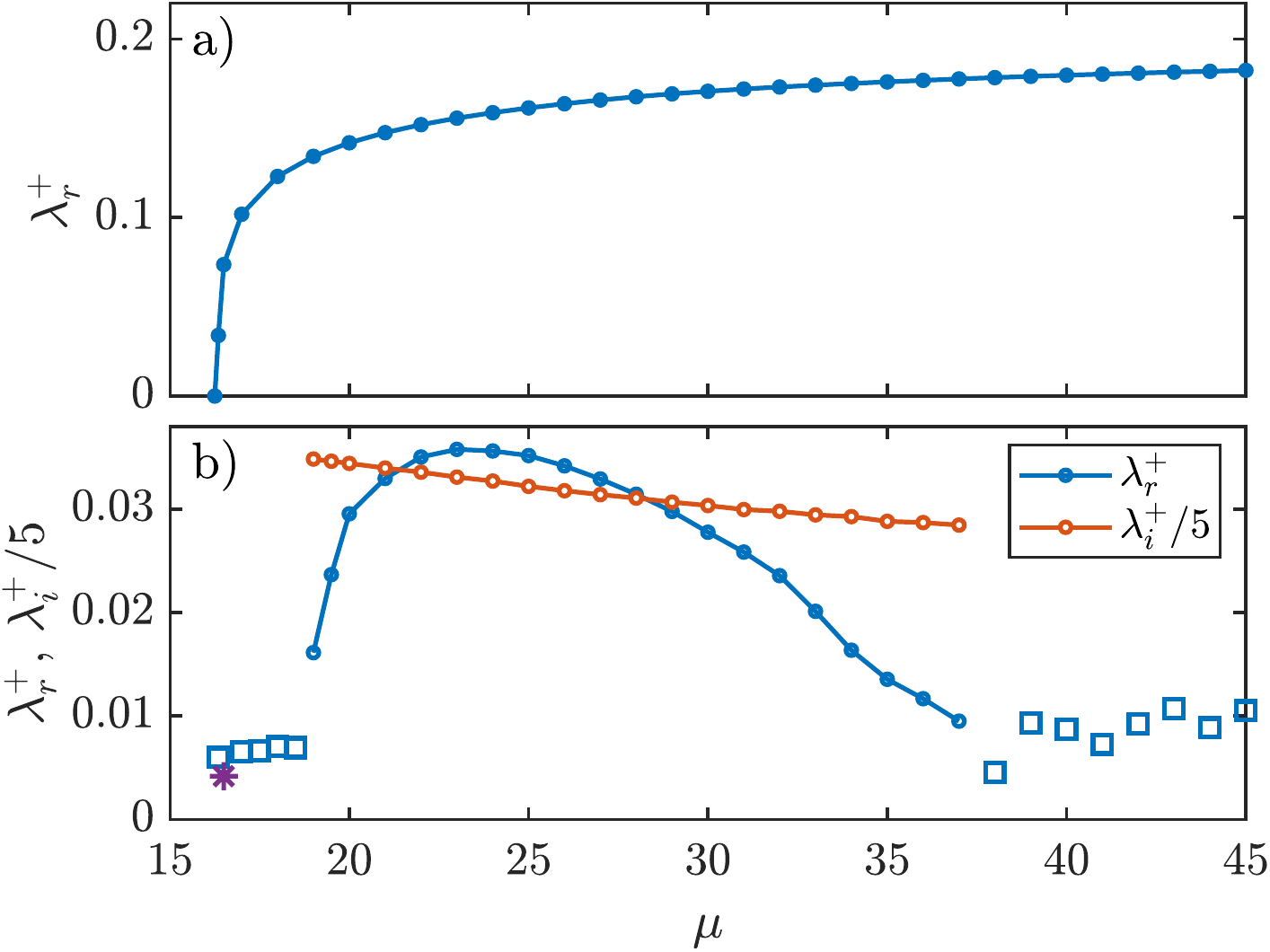}
\caption{
\label{fig:alice_real_eval}
(Color online)
Dominant (most unstable) eigenvalues for the first (top panel) and
second (bottom panel) AR families as a function of $\mu$.
(Blue) Dots and squares correspond to the real part of the eigenvalue,
while the empty (orange) dots correspond to the imaginary part of the
eigenvalue (rescaled by a factor of 5).
The first AR family always exhibits a purely real eigenvalue
indicating a dominant exponential instability.
For the second AR,
both, the small (blue) dots and (blue) squares, correspond to numerics
performed with a grid of 121$\times$121$\times$121 mesh points.
The (blue) dots correspond to an oscillatory instability while
the (blue) squares correspond to an apparently spurious set of exponential
instabilities resulting from the discretization effects in the numerics.
The (purple) asterisk corresponds to more refined numerics using
a grid of 153$\times$153$\times$153 mesh points.
}
\end{figure}
%%%%%%%%%%%%%%%%%%%%%%%%%%%%%%%%%%%%%%%%%%%%%%%%%%%%%%%%%%%%%%%

%%%%%%%%%%%%%%%%%%%%%%%%%%%%%%%%%%%%%%%%%%%%%%%%%%%%%%%%%%%%%%%
\subsection{Stability}
%%%%%%%%%%%%%%%%%%%%%%%%%%%%%%%%%%%%%%%%%%%%%%%%%%%%%%%%%%%%%%%

Let us now study the stability properties of the AR solution.
Figure~\ref{fig:alice_mu20_evals} depicts a typical
most unstable eigenvector for the first AR solution for $\mu=20$.
As for the monopole, this AR solution is unstable with a positive
real eigenvalue corresponding to an exponential instability.
As far as the eigenvector is concerned, a similar conclusion as with the
corresponding one for the monopole ensues. Namely, the eigenvector
bears a different symmetry structure than the vortex line solutions in the
$\psi_{-1}$ and $\psi_{+1}$ components, and, thus, when
added to them induces a (larger) separation between the vortex lines.
Figure~\ref{fig:alice_real_eval} depicts the instability rate for
both the first (top panel) and the second (bottom panel)
AR families as a function of $\mu$.
As mentioned above, both AR families emerge around $\mu=16.35$
in a saddle-center  bifurcation as suggested by the figure.
The first AR is clearly unstable bearing an exponential instability
over the range of chemical potentials studied.
On the other hand, in the case of the second AR shown in the bottom
panel, extrapolation of the eigenvalues for even finer
meshes suggests that the eigenvalues depicted by the (blue) squares
are spurious. In fact, as detailed in the Appendix \ref{app:numB}
(see Fig.~\ref{fig:eval_vs_N_AR2}), as the number of mesh points
($N \thintimes N \thintimes N$) increases, these (spurious) eigenvalues
tend to zero according to the power law $\propto 1/N$.
This strongly suggests that the second AR is indeed spectrally
stable in the region $16.35\leq\mu\lesssim19$ and $\mu\gtrsim38$;
namely before and after it incurs a bubble of oscillatory
instabilities (for which we can also identify and show within the
panel the relevant imaginary part).
Furthermore, not only does the extrapolation of the eigenvalues for finer
meshes suggest that the second AR is stable, but this is also in line
with the generic
bifurcation-theory-based expectation  for a saddle-center
bifurcation where the second AR is the stable
sibling of the first (unstable) AR at the bifurcation point.

As the chemical potential is further increased,
while the first AR branch retains its exponential
instability mode, the second branch AR solutions transition,
around $\mu\approx 19$, to an instability with a dominant complex
eigenvalue quartet indicating an oscillatory instability.
Finally, as $\mu$ is increased further ($\mu\gtrsim 38$), the oscillatory
instability recedes and the second AR recovers its stability.
This oscillatory instability bubble (window) for $19\lesssim\mu\lesssim 38$ emerges
when two pairs of opposite Krein signature~\cite{frantzeskakis2015defocusing}
eigenvalues collide, on the
imaginary axis, and eject as a quartet with non-zero real parts, in
line with our earlier discussion of such features in the vicinity of
the linear limit for the case of the monopole. 

%%%%%%%%%%%%%%%%%%%%%%%%%%%%%%%%%%%%%%%%%%%%%%%%%%%%%%%%%%%%%%%
\begin{figure}[!htbp]
\includegraphics[height=4.3cm]{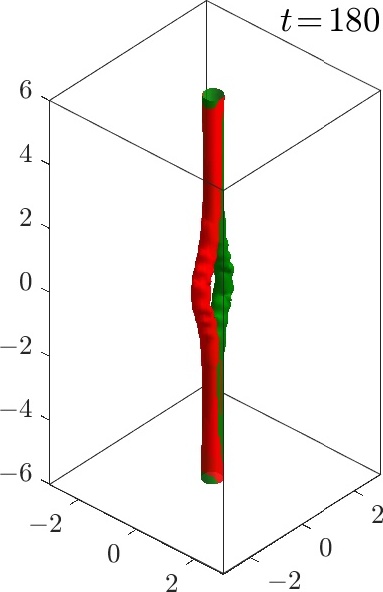}
\includegraphics[height=4.3cm]{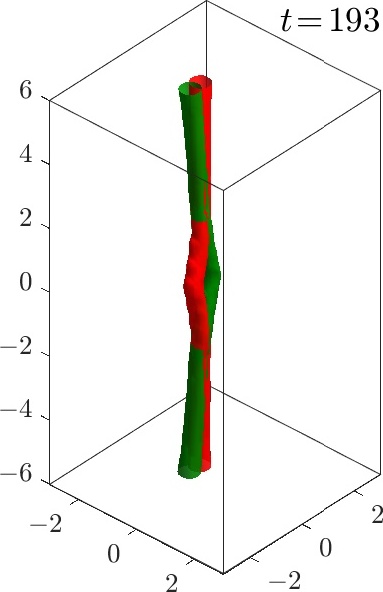}
\includegraphics[height=4.3cm]{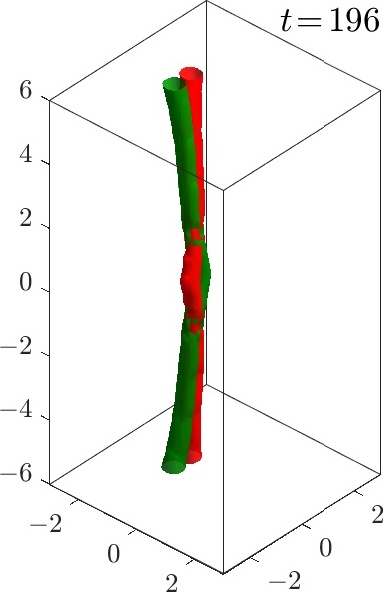}
\includegraphics[height=4.3cm]{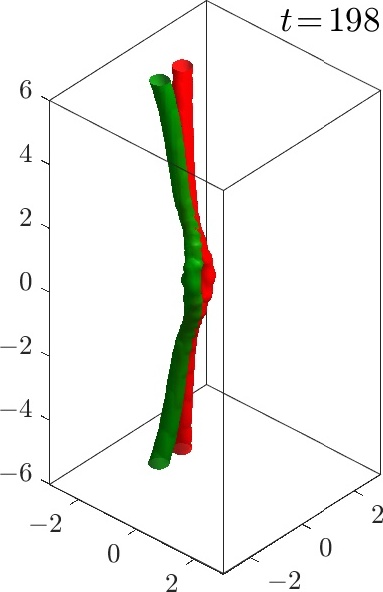}
\includegraphics[height=4.3cm]{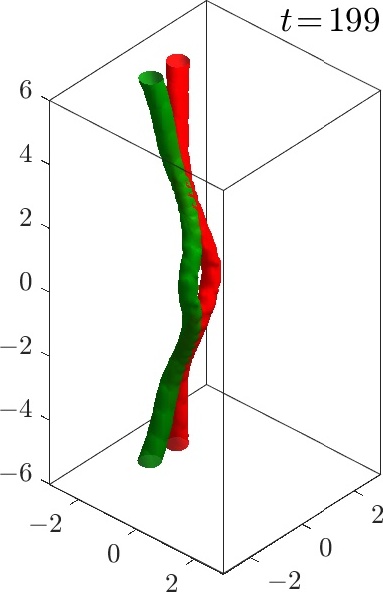}
\includegraphics[height=4.3cm]{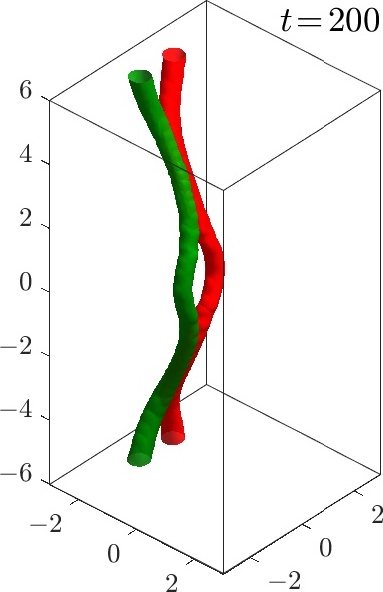}
\caption{
\label{fig:alice_mu25_dyn}
(Color online)
Destabilization dynamics for the first AR for $\mu=25$.
Same layout and perturbation as in Fig.~\ref{fig:monopole_mu20_dyn}.
For the corresponding movie please follow this
\href{https://drive.google.com/file/d/1QyW4ORZ8aJbOPUEwhZbtOOdUpkbWVX6t}{link}.
}
\end{figure}
%%%%%%%%%%%%%%%%%%%%%%%%%%%%%%%%%%%%%%%%%%%%%%%%%%%%%%%%%%%%%%%

%%%%%%%%%%%%%%%%%%%%%%%%%%%%%%%%%%%%%%%%%%%%%%%%%%%%%%%%%%%%%%%
\begin{figure}[!htbp]
\includegraphics[height=4.3cm]{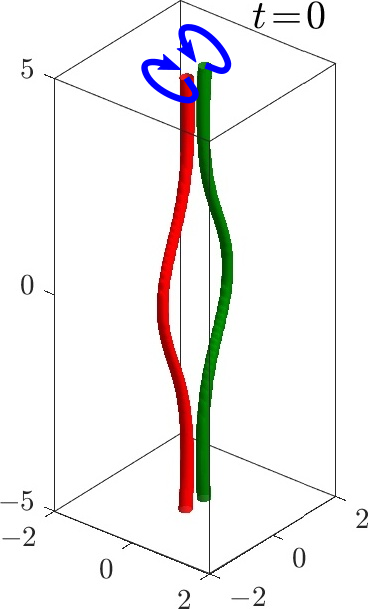}
\includegraphics[height=4.3cm]{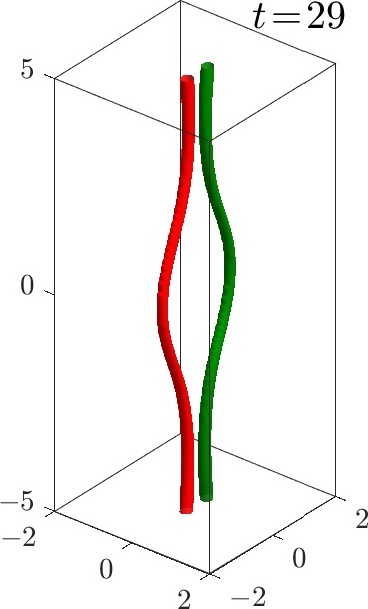}
\includegraphics[height=4.3cm]{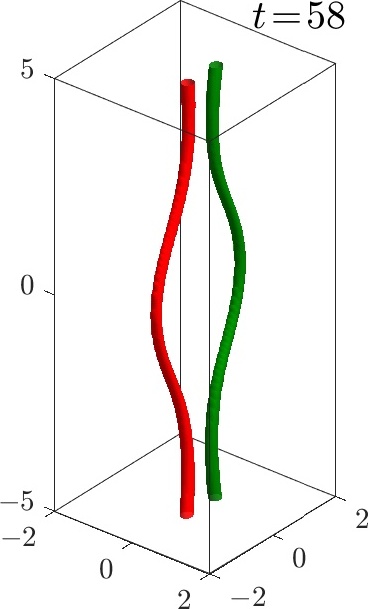}
\includegraphics[height=4.3cm]{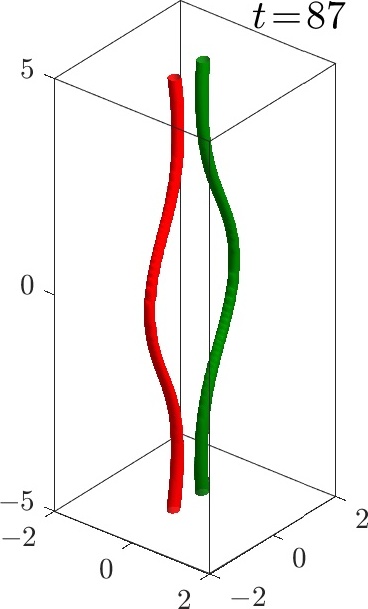}
\includegraphics[height=4.3cm]{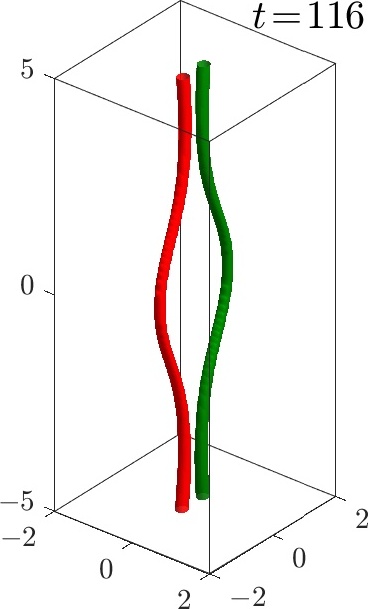}
\includegraphics[height=4.3cm]{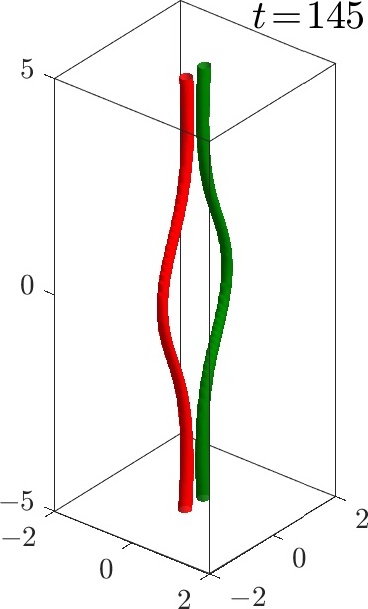}
\caption{
\label{fig:alice_homoclinic}
(Color online)
Destabilization dynamics for the first AR for $\mu=16.5$ giving rise
to a homoclinic orbit.
The vortex filaments move in a cycle depicted by the blue arrows in the
first panel.
Same layout and perturbation as in Fig.~\ref{fig:monopole_mu20_dyn}.
For the corresponding movie please follow this
\href{https://drive.google.com/file/d/1F3vihz8VVxyVq31NEjeJenISfiysuV0D}{link}.
}
\end{figure}
%%%%%%%%%%%%%%%%%%%%%%%%%%%%%%%%%%%%%%%%%%%%%%%%%%%%%%%%%%%%%%%

%%%%%%%%%%%%%%%%%%%%%%%%%%%%%%%%%%%%%%%%%%%%%%%%%%%%%%%%%%%%%%%
\begin{figure}[!htbp]
\center{
\hspace{-0.5cm}
 \includegraphics[width=0.99\columnwidth]{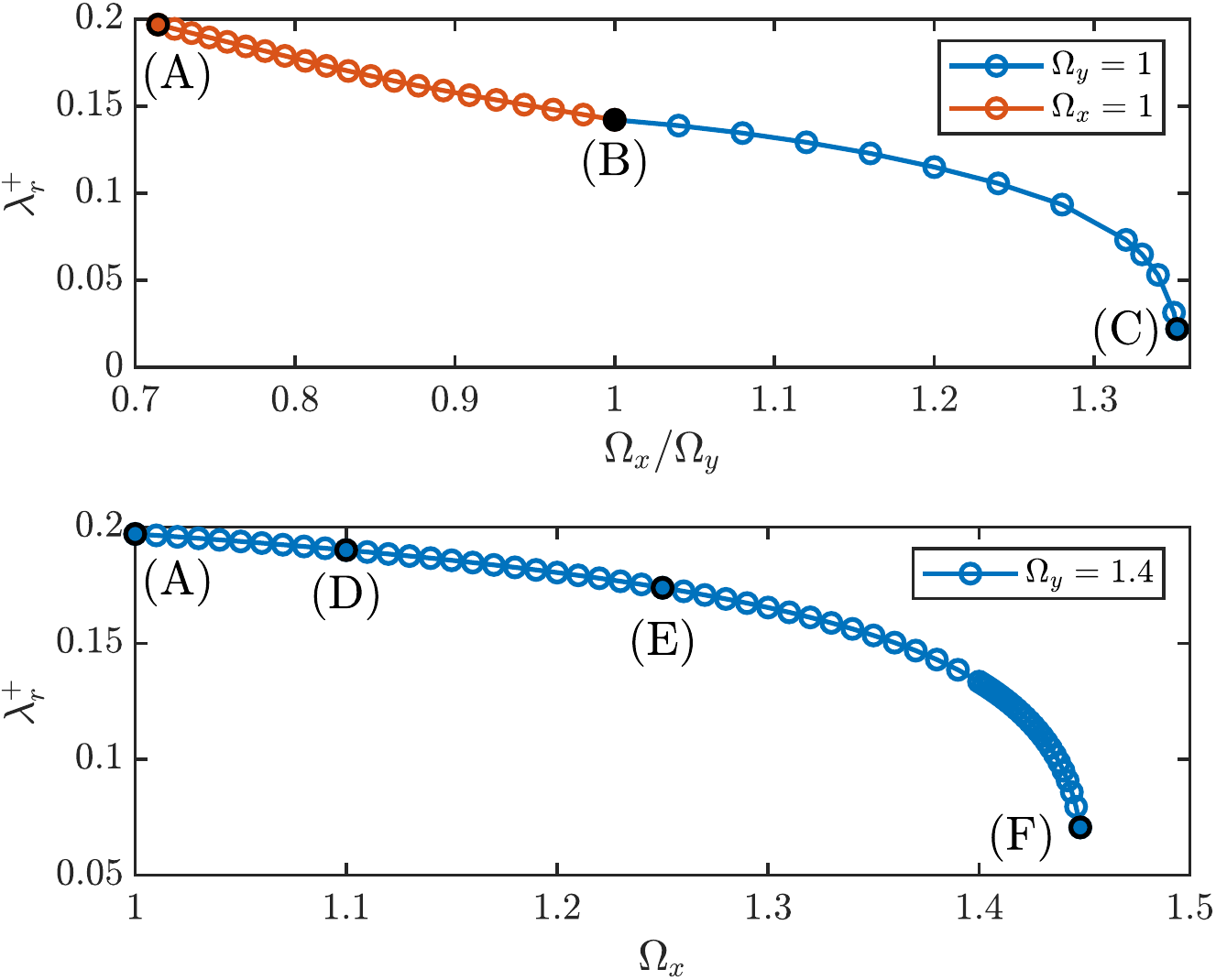}
\\[2.0ex]
 \includegraphics[width=0.99\columnwidth]{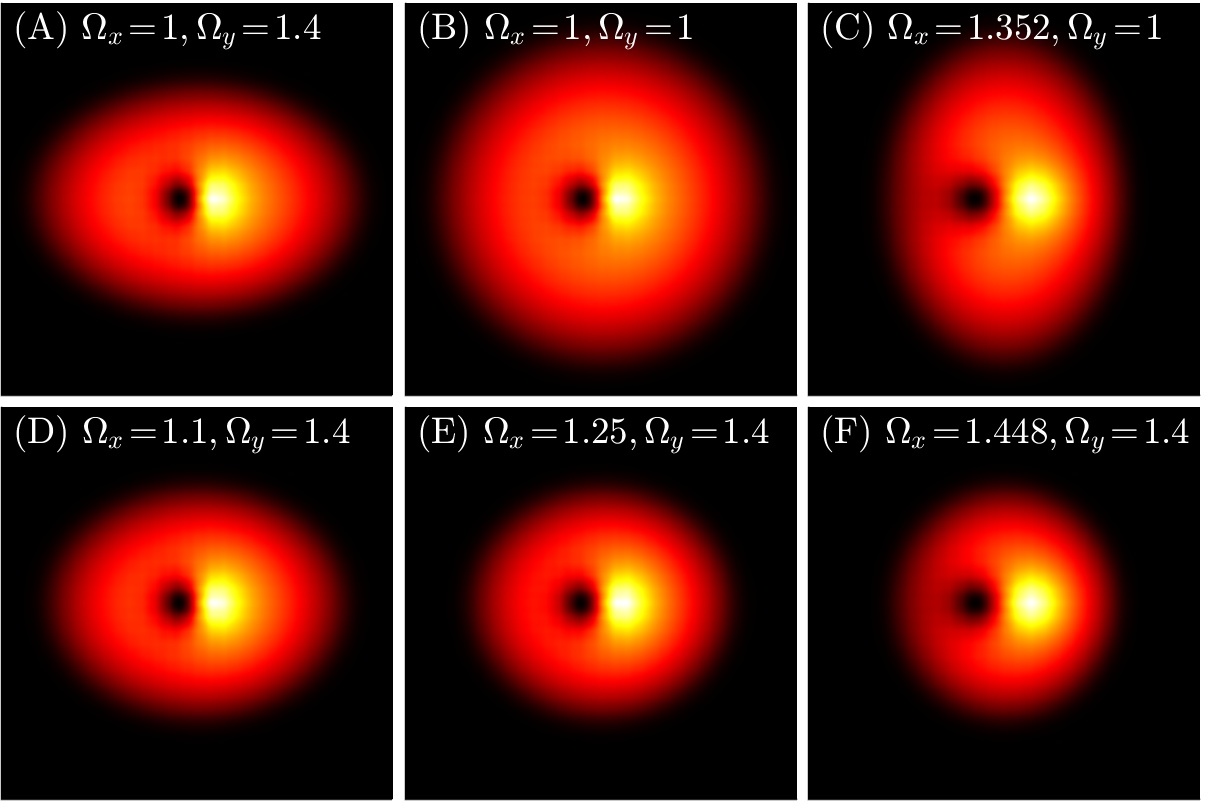}
}
\caption{
\label{fig:spectrum_alice_OR}
(Color online)
Most unstable eigenvalue (purely real) for the first AR solution
for $\mu=20$ and several trap strength combinations.
Top:
To the left of the isotropic case $\Omega_x=\Omega_y=1$ (see black dot)
the trap is oblate with $\Omega_y>1$ and $\Omega_x=1$;
while to the right of the isotropic case the trap is prolate with
$\Omega_x>1$ and $\Omega_y=1$.
The solution cannot be continued for larger $\Omega_x$ values past the point (C).
Middle:
A tighter trap with $\Omega_x=1$ and $\Omega_y>1$.
The solution cannot be continued for larger $\Omega_x$ values past the point (F).
The bottom set of panels depicts the density of the $\psi_{+1}$
component at $z=0$ corresponding to the different points labelled
in the top two panels. The field of view corresponds to
$(x,y)\in [-7,7]\times[-7,7]$.
}
\end{figure}
%%%%%%%%%%%%%%%%%%%%%%%%%%%%%%%%%%%%%%%%%%%%%%%%%%%%%%%%%%%%%%%

%%%%%%%%%%%%%%%%%%%%%%%%%%%%%%%%%%%%%%%%%%%%%%%%%%%%%%%%%%%%%%%
\subsection{Dynamics}
%%%%%%%%%%%%%%%%%%%%%%%%%%%%%%%%%%%%%%%%%%%%%%%%%%%%%%%%%%%%%%%

Figure~\ref{fig:alice_mu25_dyn} depicts the destabilization for the
first AR for $\mu=25$. As mentioned above, the destabilization from
the most unstable eigenvector tends to displace the vortex lines in
the $\psi_{-1}$ and $\psi_{+1}$ components. As Fig.~\ref{fig:alice_mu25_dyn}
shows, this displacement tends to shrink the diameter of the AR,
and at the same time tends to separate the (previously coincident)
ends of the vortex lines
across components. As a result of the above rearrangements,
the AR acquires an effective non-zero velocity and starts to advance towards $y>0$ while
the vortex lines, being reversed with respect to each other when compared to
the half ring on each side of the AR (see the figure), tend to move in the
opposite ($y<0$) direction.
As a result the vortex lines develop a strong undulation that eventually
has the two vortex lines perform complex interactions (results not shown here).
A particularly pronounced feature of these interactions is the separation/deviation
from the aligned state of the vortex filaments (of the $\psi_{\pm 1}$ components)
in the far field.

Finally, in Fig.~\ref{fig:alice_homoclinic} we depict an interesting case for
the destabilization dynamics of a first AR for $\mu=16.5$ where, after doing
a circular excursion away from the origin, the vortex filaments come back
(very close) to their original location. This type of dynamics
suggests the
possibility for the existence of homoclinic as well as
{periodic} orbits with non-trivial vortex
filament dynamics.
Indeed, this is at the heart of the characterization of
this bifurcation as a saddle-center one: the dynamics apparently
departs along the unstable manifold of the saddle and performs
an oscillation around the stable center (i.e., the second AR),
before nearly returning along the stable manifold of the
saddle. Of course, the full PDE dynamics is more complex than this
simple normal-form-based picture, yet for
substantially long times, the PDE closely reflects this effectively
one degree-of-freedom perspective.
Further time integration of the evolution presented in
Fig.~\ref{fig:alice_homoclinic} tends to destabilize the filaments in a similar
manner as the interactions depicted in
Fig.~\ref{fig:alice_mu25_dyn}.

%%%%%%%%%%%%%%%%%%%%%%%%%%%%%%%%%%%%%%%%%%%%%%%%%%%%%%%%%%%%%%%
\subsection{Towards Stabilization of the First Alice Ring}
\label{sec:AR_stab}
%%%%%%%%%%%%%%%%%%%%%%%%%%%%%%%%%%%%%%%%%%%%%%%%%%%%%%%%%%%%%%%

Finally, we briefly summarize our attempts to stabilize the first AR
solution which was found to be unstable for all the parameter
values that we used in the case of an isotropic trap.
Note that we did not try to study the effects of stabilization
for the second AR as this solution already has a window of
stability close to its bifurcation inception.
The idea to tame the instability of the first AR relies on using
the flexibility given by manipulating
the trapping strengths along the different spatial directions to create
an anisotropic trapping potential that might suppress the destabilization
along a particular direction.
For instance, Fig.~\ref{fig:spectrum_alice_OR} depicts our attempts to
stabilize the first AR by keeping $\Omega_z=1$ and varying $\Omega_x$
and $\Omega_y$. The top panel shows the most unstable (real) eigenvalue
along two different solution branches. The left (orange) branch corresponds
to an oblate trap with $\Omega_x=1$ and $\Omega_y>1$, while the right (blue)
branch corresponds to a prolate trap with $\Omega_y=1$ and $\Omega_x>1$;
see the corresponding $\psi_{+1}$ cuts labelled (A) oblate, (B) isotropic, and
(C) prolate in the lower panels of the figure.
These results tend to suggest that a prolate trap with $\Omega_y=1$ and $\Omega_x>1$
is able to attenuate the instability of the first AR. Unfortunately, this
AR solution cannot be continued past $\Omega_x\approx 1.352$ [see point (C)].
Nonetheless, a noticeable reduction of the instability growth rate can be achieved in
this prolate limit.
Finally, encouraged by these stabilization results, we also tried a tighter trap
by letting $\Omega_y=1.4$ and increasing $\Omega_x$. The stabilization results are
presented in the second panel of Fig.~\ref{fig:spectrum_alice_OR}. In this case
we again see a tendency towards stabilization as $\Omega_x$ is increased. However, as
before, the solution ceases to exist for larger $\Omega_x$ values
[$\Omega_x\approx 1.448$; see point (F)] before complete stabilization is achieved.
In short, we have been able to use the manipulation of the trap strengths
towards the reduction of the instability growth rate
of the first AR solution. Nonetheless, we have not been able to fully suppress
the relevant instability of this AR through such trap variations.
It is worthwhile to note that the work of Ref.~\cite{Ruostekoski_2003}
claims the complete stabilization of their AR
via a pair of blue-detuned focused Gaussian laser beams.
However, in the study herein, our focus was to study
only the role of the most ubiquitous magnetic trap setting rather than
towards the inclusion of additional forms of confinement.

%%%%%%%%%%%%%%%%%%%%%%%%%%%%%%%%%%%%%%%%%%%%%%%%%%%%%%%%%%%%%%%
\section{Conclusions and Future Work}
\label{sec:conclu}
%%%%%%%%%%%%%%%%%%%%%%%%%%%%%%%%%%%%%%%%%%%%%%%%%%%%%%%%%%%%%%%

In the present work, we explored the existence, stability, and dynamics
of monopole and Alice ring (AR) solutions in a polar (anti-ferromagnetic)
$F=1$ spinor condensate.
Our study was primarily numerical in nature but it was critically
assisted in the obtained understanding by theoretical limits such as
the linear and the large density (Thomas-Fermi) ones,
and notions such as those of the director vector and the magnetic
quadrupolar tensor, the BdG spectral analysis and the evolution of
the vorticity field among others.
The monopole solution was found to emanate from the linear (low density)
limit as the combination of opposite sign, overlapping, vortex lines
in the $\psi_+$ and $\psi_-$ components and a domain wall (bearing a
planar phase jump) in the $\psi_0$ component.
The monopole solution was found to be unstable for all chemical
potentials that we tested through a pair of real eigenvalues
associated with an exponential instability.
Interestingly, for large enough chemical potentials, the instability
of the monopole typically forms, in the first time interval of its destabilization,
a transient profile where the two vortex lines bulge locally near the center
creating a ring completed by two halves corresponding to each vortex
line across the $\psi_{\pm 1}$ components.

Motivated by such transient waveforms from the monopole destabilization
and leveraging such a singular ring profile into a fixed point iteration procedure
revealed the existence of a stationary ring structure. By analyzing the corresponding
nematic vector we corroborated the fact that this stationary state corresponds to an
AR whose topological texture in the far field matches the monopole
(radial outward field), yet it contains a $\pi$-disclination of the
nematic phase vector across the ring.
Continuation analysis over the chemical potential reveals that the
AR comes in two sizes: a smaller and a larger AR.
These two AR families are found to collide in (i.e., bifurcate through)  a
saddle-center scenario at a critical value of the chemical potential,
and the two solutions co-exist past this turning point as the chemical
potential is increased.
The smaller AR was found to always be unstable with a purely real
(exponential) instability.
In contrast, although finite-size numerics display a small instability for
the larger ring, extrapolation of the associated eigenvalues for finer and finer
meshes reveal a power-law decay for this instability. This is strongly suggestive
that these small instabilities are spurious and induced by the mesh discretization
and that the larger ring is indeed {\em spectrally stable}  close to its
bifurcation with the other ring. The larger ring is stable from
its inception until it destabilizes for a chemical potential window
(or bubble) with an oscillatory instability.
As the chemical potential is increased further, this oscillatory instability
disappears and the larger AR seems to regain its stability.

In an attempt to stabilize the smaller AR we probed its stability when
the trapping deviated from its isotropic (spherical) starting point.
Our results suggest that a prolate trap is able to attenuate the instability
for the smaller AR but, unfortunately, is
not enough to render it stable in the parameter regimes that we explored.
Nonetheless, this noticeable attenuation of the instability might
facilitate the observability of the smaller AR in addition to the
larger AR which is stable for suitable values of the chemical potential.

It is, of course,
relevant to note here that experiments such as those of Ref.~\cite{dsh1}
for monopoles and also, e.g., those of Refs.~\cite{dsh2,dsh3} for other
complex topological patterns such as quantum knots strongly suggest that
the configurations presented herein are possible to realize in the
current experimental state-of-the-art. Moreover, one can think of
numerous directions of potential extensions of the present work to more
complex settings. On the one hand, including additional optical
potentials and exploring modifications to the branches of solutions presented herein
(and, importantly, their stability) is an important future step, in
line also with the original suggestion of Ref.~\cite{Ruostekoski_2003}.
On the other hand, and perhaps even more challengingly, one can
consider extensions of the present setting into the 5-component setting
of $F=2$~\cite{KAWAGUCHI2012253}. While it is natural to consider
extensions of the monopole state in the latter setting, notions
such as the one of the AR are far more elusive for $F=2$
and to the best of our knowledge have never been systematically
examined in either numerical or physical experiments. Such studies
are currently in progress and will be reported in future publications.

\section*{Acknowledgments}
We gratefully acknowledge fruitful discussions with Mikko M\"ott\"onen.
This material is based upon work supported by the US
National Science Foundation under Grants No.~PHY-1603058 and
PHY-2110038 (R.C.G.), No.~PHY-1806318 (D.S.H.), and No. PHY-2110030 (P.G.K.).

\appendix

%%%%%%%%%%%%%%%%%%%%%%%%%%%%%%%%%%%%%%%%%%%%%%%%%%%%%%%%%%%%%%%%%%%%%%%%%%%%%
\section{Linearized GP equations}
\label{app:BdG}
%%%%%%%%%%%%%%%%%%%%%%%%%%%%%%%%%%%%%%%%%%%%%%%%%%%%%%%%%%%%%%%%%%%%%%%%%%%%%

The adimensionalized GP equations [cf. Eqs.~\eqref{eq:GPd}] can be cast in
the matrix form
\begin{equation}
  i\frac{\partial\Psi}{\partial t}= \left(\mathcal{H} +c_2 A \right) \Psi,
\end{equation}
where
\[
\mathcal{H}=-\frac{1}{2}\nabla^2+V(\textbf{r})+c_0 n,
\]
and
\[
A=
  \begin{bmatrix}
    n_{+1}+n_0-n_{-1} &        0   &   0  \\
         0         & n_{+1}+n_{-1} &   0  \\
         0         &     0      &  n_{-1}+n_0-n_{+1}
  \end{bmatrix}
,
\]
where we recall that $n_i=|\psi_i|^2=\psi_i^{\ast}\psi_i$.
Then, by following the corresponding evolution about the steady state
$\Psi_0=(\psi_{+1},\psi_0,\psi_{-1})$, the solution
$\Psi=\Psi_0+\varepsilon\,\delta\Psi$ yields the
linearization ($\varepsilon\ll1$) dynamics
\begin{equation}
  i\frac{\partial\delta \Psi}{\partial t}= K_R\, \delta\Psi+ K_I\, \delta\Psi^{\ast},
\end{equation}
where
\begin{eqnarray}
  K_R&=&\mathcal{H}+ c_0 B +c_2 (C+D),
\notag
\\
\notag
  K_I&=&\mathcal{H}+c_0 E+ c_2(F+ G),
\end{eqnarray}
and
\[
B=
  \begin{bmatrix}
     n_{+1} &  \psi_0^{\ast}\psi_{+1}  &  \psi_{-1}^{\ast}\psi_{+1}  \\
     \psi_{+1}^{\ast}\psi_0 &  n_0&  \psi_{-1}^{\ast}\psi_0\\
     \psi_{+1}^{\ast}\psi_{-1} &  \psi_{0}^{\ast}\psi_{-1} &  n_{-1}
  \end{bmatrix}
,
\]
\[
C=
  \begin{bmatrix}
    F_z+n_{+1}+n_0 & \psi_{0}^{\ast}\psi_{+1} & -\psi_{-1}^{\ast}\psi_{+1} \\
    \psi_{+1}^{\ast}\psi_{0} & n_{-1}+n_{+1}& \psi_{-1}^{\ast}\psi_{0} \\
   - \psi_{+1}^{\ast}\psi_{-1}  & \psi_{0}^{\ast}\psi_{-1}  & F_z+n_{-1}+n_0
  \end{bmatrix}
,
\]
\[
D=
  \begin{bmatrix}
    0 & 2 \psi_{-1}^{\ast}\psi_0 &0  \\
   2 \psi_0^{\ast}\psi_{-1} & 0 & 2 \psi_0^{\ast}\psi_{+1}\\
    0 & 2\psi_{+1}^{\ast}\psi_0 & 0
  \end{bmatrix}
,
\]
\[
E=
  \begin{bmatrix}
    \psi_{+1}^{2} & \psi_{0} \psi_{+1} & \psi_{-1}\psi_{+1} \\
    \psi_{+1} \psi_{0} &  \psi_{0}^2 & \psi_{-1} \psi_{0} \\
    \psi_{+1} \psi_{-1}  & \psi_{0} \psi_{-1}  & \psi_{-1}^{2}
  \end{bmatrix}
,
\]
\[
F=
  \begin{bmatrix}
    \psi_{+1}^{2} & \psi_{0} \psi_{+1} & -\psi_{-1}\psi_{+1} \\
    \psi_{+1} \psi_{0} &  0 & \psi_{-1} \psi_{0} \\
   -\psi_{+1} \psi_{-1}  & \psi_{0} \psi_{-1}  & \psi_{-1}^{2}
  \end{bmatrix}
,
\]
\[
G=
  \begin{bmatrix}
   0 & 0 &\psi_{0}^{2}  \\
   0 & 2 \psi_{+1}\psi_{-1} & 0\\
   \psi_{0}^{2} & 0 & 0
  \end{bmatrix}
.
\]
Finally, for a perturbation of the form
$\delta \Psi = P e^{\lambda t}+ Q^{\ast} e^{\lambda^{\ast} t}$, the linearized
problem takes the form
\begin{equation}
\notag
\left\{
\begin{array}{rcl}
i\lambda P &=& K_R P +K_I Q\\[1.0ex]
i\lambda Q &=& - K_R^{\ast} Q - K_I^{\ast} P
\end{array}
\right.
\end{equation}
which can be explicitly written as the eigenvalue problem
\begin{equation}
\label{eq:evalpb}
M\mathcal{V} = \lambda \mathcal{V},
\end{equation}
with
\begin{equation}
\notag
M=
-i
\begin{bmatrix}
K_R & K_I\\
-K_I^{\ast} & -K_R^{\ast}
\end{bmatrix}
\quad
\text{and}
\quad
\mathcal{V}=
\begin{bmatrix}
P\\
Q
\end{bmatrix}.
\end{equation}

%%%%%%%%%%%%%%%%%%%%%%%%%%%%%%%%%%%%%%%%%%%%%%%%%%%%%%%%%%%%%%%%%%%%%%%%%%%%%
\section{Numerics}
\label{app:num}
%%%%%%%%%%%%%%%%%%%%%%%%%%%%%%%%%%%%%%%%%%%%%%%%%%%%%%%%%%%%%%%%%%%%%%%%%%%%%

In this section, we provide
some details on the numerical challenges presented
by the system at hand and the methodologies that we opted to use.

%%%%%%%%%%%%%%%%%%%%%%%%%%%%%%%%%%%%%%%%%%%%%%%%%%%%%%%%%%%%%%
\subsection{Space and time discretization: steady states and forward integration}
\label{app:numA}
%%%%%%%%%%%%%%%%%%%%%%%%%%%%%%%%%%%%%%%%%%%%%%%%%%%%%%%%%%%%%%

We used standard, second-order accurate, finite difference (FD) discretization in space to
describe steady states, evolution solutions, and eigenfunctions. To find steady states
we discretized Eq.~(\ref{eq:GPdSS}) using FDs and used the nonlinear Newton-Krylov
solver {\tt nsoli}~\cite{NSOLI_BOOK} until a maximum residual of $10^{-15}$
was achieved. Lack of convergence below a maximum residual of $10^{-14}$ using
numerical continuation along the chemical potential $\mu$ (with steps as
small as $\Delta\mu=0.001$) was considered as indication of the termination of
the solution branch that was followed.
The dynamical evolution of the configurations that we followed was performed with
FD in space and standard fourth-order Runge-Kutta (RK4) method in
time with a (fixed) time step adjusted to avoid numerical instabilities~\cite{Caplan:13}.
The spatial mesh is defined, for the isotropic trapping $\Omega_x=\Omega_y=\Omega_z$
case, as an $N^3$ cube with homogeneous spatial spacing $dx=dy=dz$ along all
directions. For the anisotropic cases presented in Sec.~\ref{sec:AR_stab} we
kept a constant homogeneous discretization $dx=dy=dz$ but adjusted the domain
to avoid ``wasting'' mesh points by choosing a domain
$(x,y,z)\in[-L_x,L_x]\times[-L_y,L_y]\times[-L_z,L_z]$ with $L_i$ adjusted
to the closest $dx$ to fit 1.5 times the Thomas-Fermi radius in that direction.
%

%%%%%%%%%%%%%%%%%%%%%%%%%%%%%%%%%%%%%%%%%%%%%%%%%%%%%%%%%%%%%%%
\begin{figure}[!htbp]
 \includegraphics[width=0.99\columnwidth]{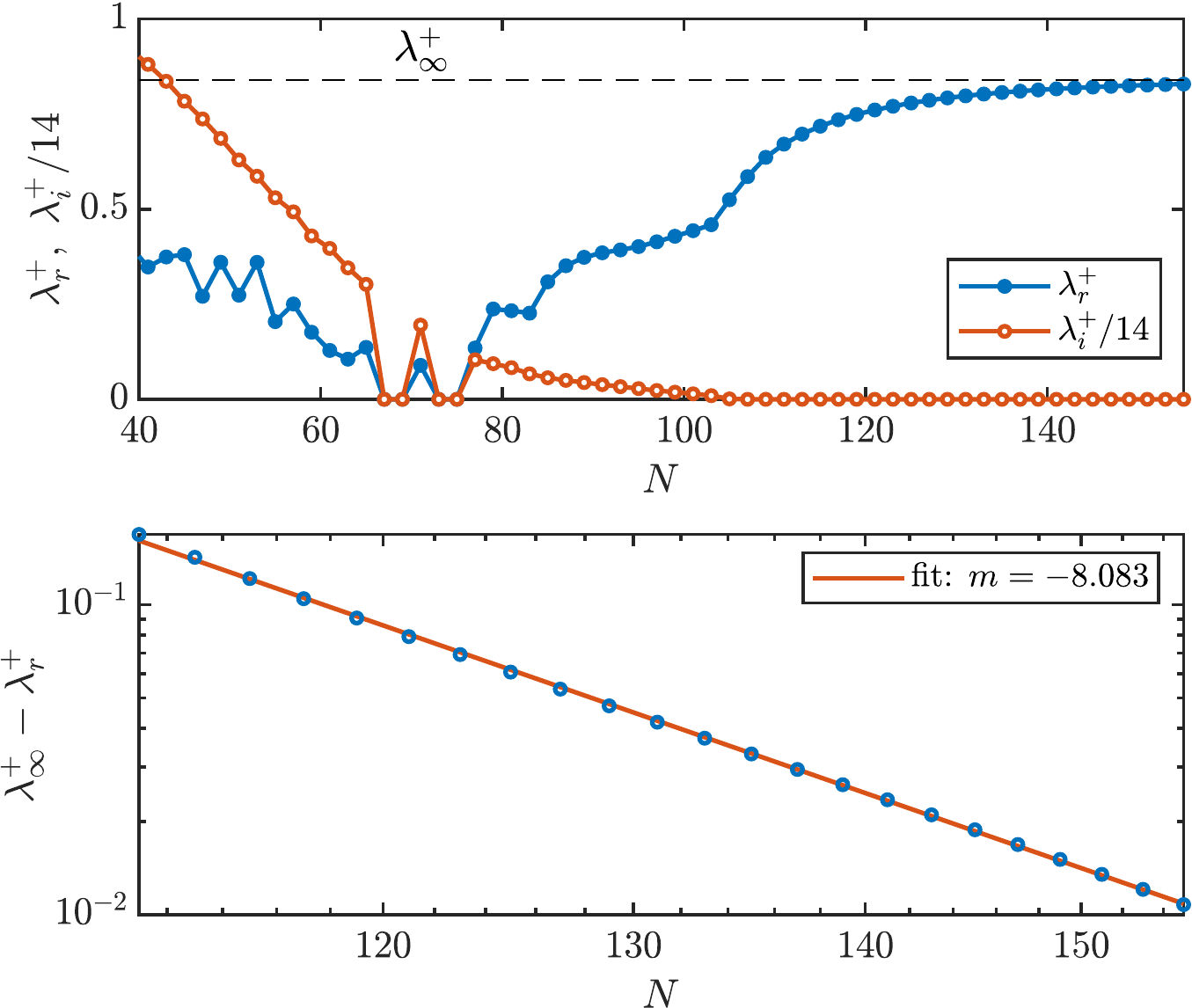}
 \caption{
\label{fig:eval_vs_N}
(Color online)
Convergence of the extracted real part of the most unstable eigenvalue
for the monopole solution for $\mu=20$ as the number of spatial mesh points
in increased.
We used a domain $(x,y,z)\in[-L_x,L_x]\times[-L_y,L_y]\times[-L_z,L_z]$
where $L_x=L_y=L_z=12$ and discretized it homogeneously along all
directions ($dx=dy=dz$) using $N$ points along each direction.
The bottom panel depicts the corresponding log-log plot of the data
([blue] circles) for larger values of $N$ and its fit ([orange] line)
suggesting that the largest unstable eigenvalue tends to
$\lambda^+_\infty\approx 0.84$ at a rate approximately $\propto 1/N^8$.
}
\end{figure}
%%%%%%%%%%%%%%%%%%%%%%%%%%%%%%%%%%%%%%%%%%%%%%%%%%%%%%%%%%%%%%%

%%%%%%%%%%%%%%%%%%%%%%%%%%%%%%%%%%%%%%%%%%%%%%%%%%%%%%%%%%%%%%%
\begin{figure}[!htbp]
 \includegraphics[width=0.99\columnwidth]{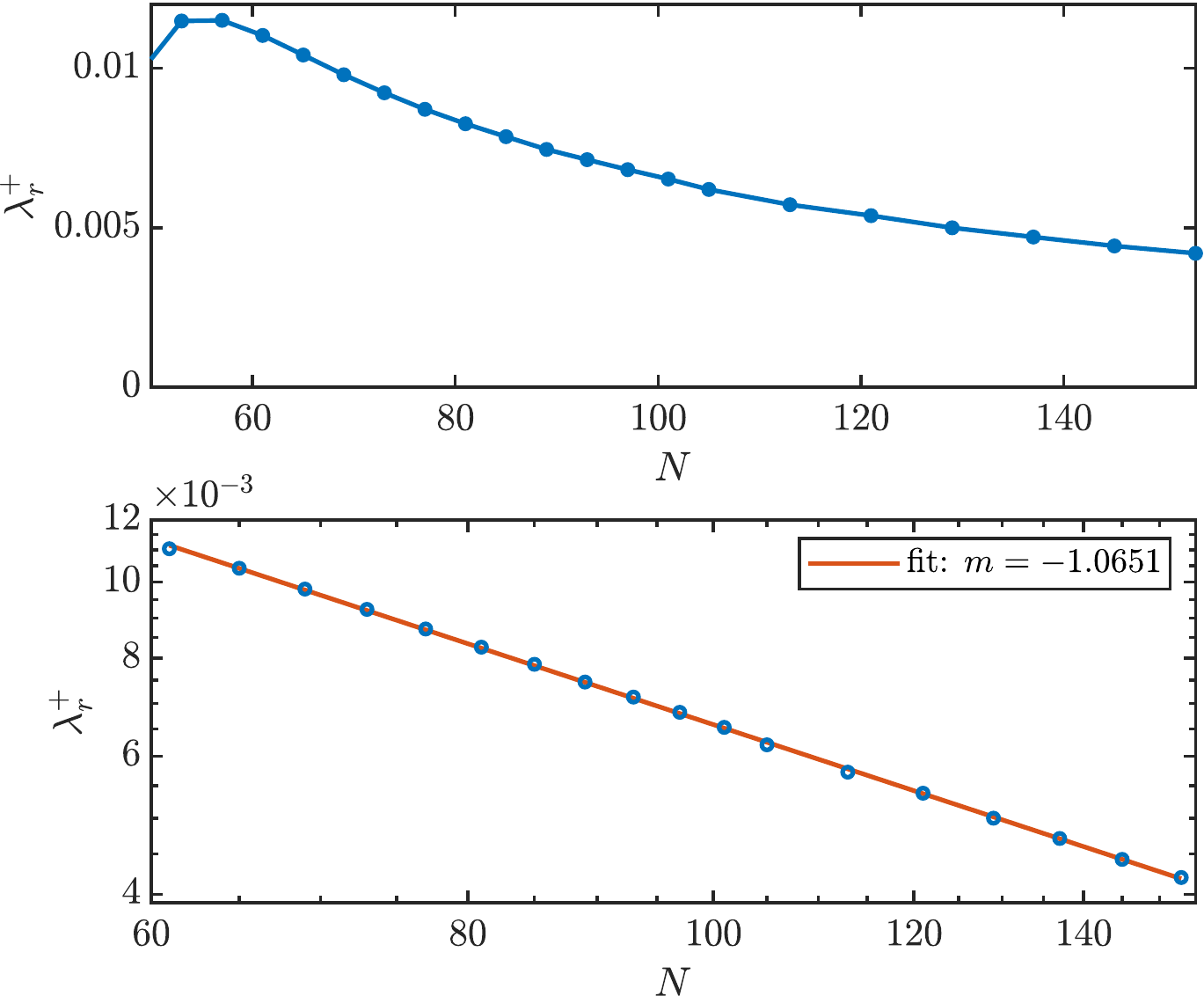}
\caption{
\label{fig:eval_vs_N_AR2}
(Color online)
Convergence of the numerically (dis\-cre\-ti\-za\-tion-based)
induced instability for the second AR for $\mu=16.5$ as the
number of mesh points $N \thintimes N \thintimes N$ is increased.
In the top panel we depict the largest real part computed from
fitting the growth of our FDs+RK4 integration over 1000
(adimensionalized) seconds.
The bottom panel depicts the corresponding log-log plot of the data
([blue] circles) for larger values of $N$ and its fit ([orange] line)
suggesting that the largest unstable eigenvalue tends to zero
at a rate approximately $\propto 1/N$.
This suggest that the second AR is indeed stable in the
continuum $N\rightarrow\infty$ limit of the actual GP model.
}
\end{figure}
%%%%%%%%%%%%%%%%%%%%%%%%%%%%%%%%%%%%%%%%%%%%%%%%%%%%%%%%%%%%%%%

%%%%%%%%%%%%%%%%%%%%%%%%%%%%%%%%%%%%%%%%%%%%%%%%%%%%%%%%%%%%%%
\subsection{Eigenvalue computations using forward integrator}
\label{app:numB}
%%%%%%%%%%%%%%%%%%%%%%%%%%%%%%%%%%%%%%%%%%%%%%%%%%%%%%%%%%%%%%

The BdG stability problem (see Appendix \ref{app:BdG}) was the most challenging
numerical aspect of this work. The idea is to discretize, using FDs, the
continuous eigenvalue problem~(\ref{eq:evalpb}). Since large eigenvalue problems
are very time and, above all, computer memory intensive, we first studied the
convergence of the most unstable mode as a function of the discretization. For this
purpose we start with the numerically computed steady state (see above) perturbed
by a random perturbation of (relative) size $10^{-10}$ and then integrate forward
using our FDs+RK4 integration scheme. By following the norm of the density difference
between the evolved state and the initial configuration, we were able to extract
(fitting using least squares) the exponential growth of perturbations and thus
extract $\lambda^{+}_r$, the real part of the most unstable eigenvalue.
Figure~\ref{fig:eval_vs_N} shows a typical convergence plot of $\lambda^{+}_r$
for the monopole solution as the number of mesh points is increased.
The bottom panel suggests that the $\lambda^{+}_r$ converges to
$\lambda^+_\infty\approx 0.84$ at a rate approximately $\propto 1/N^8$.
These results indicate that it is necessary to use a relatively large number
of mesh points to accurately capture the growth rate associated
with the most unstable eigenvalue.
Therefore, as a balance between eigenvalue convergence and a manageable system size,
we use a computational mesh with $N=129$ for the
the monopole spectra results shown in Fig.~\ref{fig:alice_real_eval}.
Given that the monopole solution is generically unstable, we find that
the above most-unstable-eigenvalue extraction is practical since, given any 
initial perturbed stationary state, the instability will eventually develop.
However, a more challenging issue arises when the solution is stable (or very
weakly unstable) as one needs to integrate for very long times to corroborate
that no instability can potentially grow. This, coupled with the issue of
needing large meshes to resolve the healing length of the vortex line,
quickly results in unmanageable numerics when extracting a BdG stability picture
as a function of one or more parameters (in our case the chemical potential $\mu$).
This problem is precisely what one faces when trying to determine the stability
of the second AR close to its emergence around $\mu=16.35$.
Figure~\ref{fig:eval_vs_N_AR2} depicts the convergence of the computed
growth rate for the second AR for $\mu=16.5$ (i.e., slightly to the right of
its emergence) as $N$ increases. The top panel indicates that the computed
eigenvalue through the FDs+RK4 integration does decrease as $N$ increases.
Similar to the convergence of the eigenvalue for the monopole
(see Fig.~\ref{fig:eval_vs_N}), the bottom panel suggests that the
second AR eigenvalue tends to {\em zero} at an approximate rate $\propto 1/N$.
Although this decay rate is weak, it does suggest that the second AR is
indeed {\em stable} at its inception around $\mu=16.35$.
This conclusion is also supported from a bifurcation viewpoint as the
first and second ARs seem to be born out of a saddle-center bifurcation
and, naturally, one branch must be unstable (first AR) and the other
{\em stable} (second AR).
In fact, we have corroborated (results not shown here) this bifurcation
scenario by following the actual BdG spectra for a smaller domain size
with $N=35$ using a combination of pseudo-arclength continuation and
the eigensolver FEAST (see below).
Therefore, we conclude that the second AR is {\em stable} in the
window $16.35\lesssim\mu\lesssim19$, namely between its emergence
and the appearance of the oscillatory instability bubble.

%%%%%%%%%%%%%%%%%%%%%%%%%%%%%%%%%%%%%%%%%%%%%%%%%%%%%%%%%%%%%%
\subsection{BdG eigenvalue and eigenvector computations using eigensolver}
\label{app:numC}
%%%%%%%%%%%%%%%%%%%%%%%%%%%%%%%%%%%%%%%%%%%%%%%%%%%%%%%%%%%%%%

It is important to mention that, due to computational limitations,
the eigenvectors shown in Figs.~\ref{fig:monopole_mu20_evals}
and~\ref{fig:alice_mu20_evals} are depicted for $N=51$ and $N=55$ for, respectively,
the monopole and AR solutions.
For the BdG eigenvalue computations, we used the FEAST eigenvalue algorithm~\cite{PhysRevB.79.115112,doi:10.1137/13090866X,doi:10.1137/15M1026572}
which combines accuracy, efficiency and robustness even for ill-conditioned
matrices (see, for example~\cite{CHARALAMPIDIS2020105255}). In the present
setup, we note that an $N^3$ mesh corresponds to a FD BdG stability matrix of
size $6N^3 \thintimes 6N^3$ as there are 3 components and the field is complex.
The eigenvalue computations were performed on an Intel(R) Core(TM) i7-8700
workstation with 64Gb of RAM. The associated FD stability matrix albeit being
highly sparse of size $795,906 \thintimes 795,906$ ($998,250 \thintimes 998,250$)
containing $9,457,236$ ($11,870,100$) non-zero elements for $N=51$ ($N=55$).
Therefore, the results depicted in Figs.~\ref{fig:monopole_mu20_evals}
and~\ref{fig:alice_mu20_evals} should be taken as {\em qualitative} rather
than quantitative and are hereby presented to help determine the
nature of the instabilities experienced by the monopole and
AR solutions.

\providecommand{\noopsort}[1]{}\providecommand{\singleletter}[1]{#1}

\end{document}